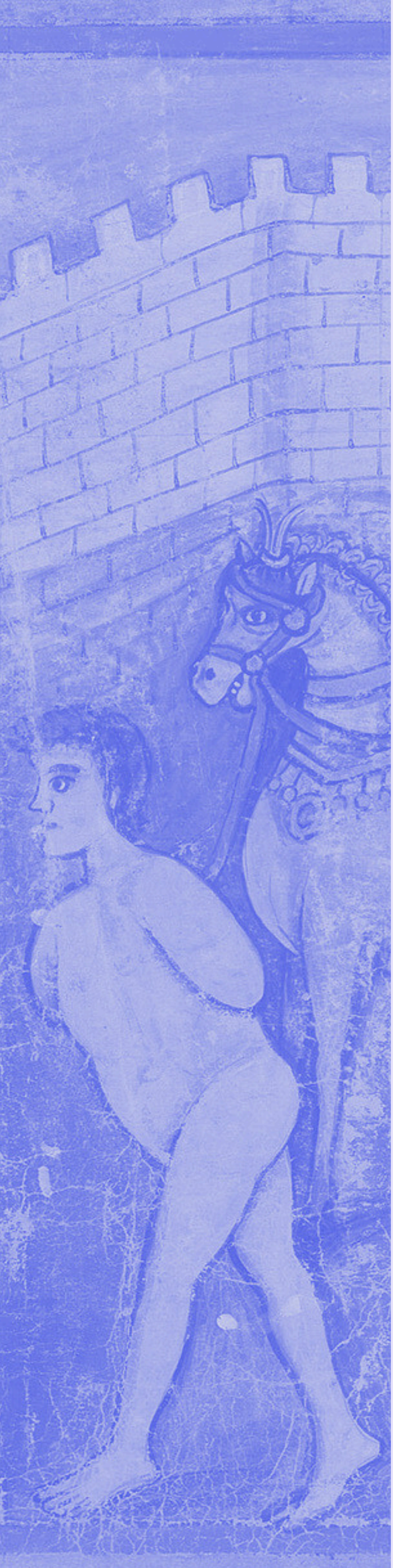

April 2026

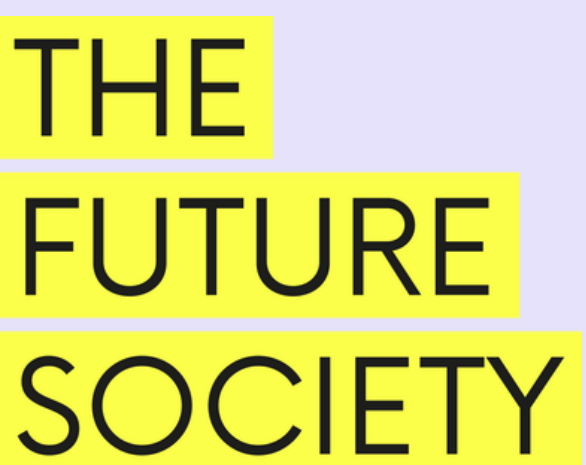

# Beware of Geeks Bearing Gifts

## Building True EU Frontier AI Sovereignty

The Future Society

**Contact:** info@thefuturesociety.org

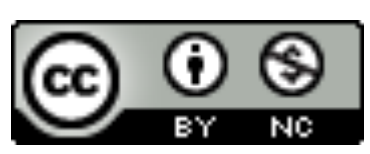

# Executive Summary

**Europe's sovereignty—its capacity to remain a self-determined political and economic actor—is at stake in the face of a new geopolitical reality.** Europe's institutional architecture, industrial base, and strategic positioning were not built for a world defined by crises and rapid technological change. COVID-19, Russia's invasion of Ukraine, and the global race over AI have made this clear in quick succession. Interests increasingly collide, and the rules that once governed international relations can no longer be taken for granted. In response, a growing body of high-level reports has called for the continent to "wake up" (von der Leyen, 2024), and the policy response is accelerating. As European Commission President von der Leyen put it at Davos in January 2026: "[T]he seismic change we are going through today is an opportunity, in fact, a necessity, to build a new form of European independence" (von der Leyen, 2026).

**Among the domains where European sovereignty will be contested, frontier AI is among the most consequential.** Unlike prior general-purpose technologies like the internet or electricity, frontier AI does not merely transmit or process but is designed to act with increasing autonomy, embedding the values, priorities, and assumptions of its developers into decision-making across economic structures, security architectures, and democratic institutions. Its impact is horizontal, projected to boost global economic output by up to 15% over the next decade (PwC, 2025). It is already deployed in military planning (Puscas, 2024), autonomous systems (ABC News, 2025), and cyber operations (Frontier Model Forum, 2024), and it is reshaping labour markets (Bengio et al., 2026), information environments, and democratic processes (Wallner et al., 2025). Yet Europe enters this transformation from a position of structural dependence. Frontier AI models come nearly exclusively from the United States or China, the US commands roughly sixteen times the EU's AI supercomputing capacity, and only 15% of global hyperscale data centre capacity is located within the EU (Pilz et al., 2025b; Synergy Research Group, 2025).

**The EU has proposed a substantial set of frontier AI policies, but these remain fragmented and present insufficient ambition to truly secure frontier AI sovereignty.** Following the 2024 Draghi Report, the European Commission has advanced initiatives spanning areas like compute infrastructure, data governance, regulatory simplification, skills, and research funding. Our analysis of 92 initiatives announced across four major Commission communications in Section 5.2 reveals that while the portfolio exhibits genuine strategic intent, these efforts often appear disconnected, lacking a cohesive vision for a sovereign frontier AI stack and focusing narrowly on short-term economic gains.

**We propose discussing frontier AI sovereignty as a means and not an end, extending the debate well beyond narrow economic competitiveness.** Our contribution is twofold: (i) making policy trade-offs more explicit and policy objectives more nuanced, moving beyond the simplistic equation of sovereignty with short-term economic competitiveness; and (ii) identifying blind spots and redundancies in industrial policy for frontier AI value chain governance. The report is structured as follows:

1. **First, we decompose sovereignty into five distinct pillars that collectively define what it means for Europe to remain a self-determined actor (Section 3).** Through a review of the sovereignty debate, we identify five objectives that recur across the literature: (i) Economic Competitiveness; (ii) Resilience and Preparedness; (iii) Security and Defence; (iv) European Values and Identity; and (v) Foreign Relations and Perceptions. Such a multi-faceted understanding advances existing sovereignty discussions, which are often narrowly focused on short-term commercial considerations.

2. **Second, we comprehensively map the frontier AI stack (Section 4).** Frontier AI does not rest on a single technology but on a complex, layered stack in which each layer builds on the one below. We decompose this stack into five layers comprising 26 components and 29 sub-components, at a finer level of granularity than most prior efforts. This decomposition enhances policymakers' capacity to identify which specific layers demand intervention, which can serve as leverage points, and where the EU holds structural strengths or critical vulnerabilities.

3. **Third, we present a unified framework connecting the five sovereignty pillars to the frontier AI stack, that enables the design, assessment, and prioritisation of frontier AI policy interventions through a comprehensive sovereignty lens (Section 5.1).** The framework elucidates trade-offs that the EU's currently scattered initiatives targeting the frontier AI stack tend to obscure—recognising that advancing one pillar can come at the direct expense of another. For instance, expanding the availability of sensitive user data may further short-term Economic Competitiveness (pillar 1) by enabling richer user profiles, but it may come at the cost of Security and Defence (pillar 3), as well as European Values (pillar 4), by increasing vulnerability to foreign influence campaigns and undermining citizens' right to privacy.

4. **Finally, we apply the framework to five EU initiatives and provide an in-depth discussion of AI Gigafactories, revealing concrete trade-offs that current policy debates leave implicit (Section 5.3–5.4).** We first illustrate the framework's application across five initiatives selected from the 92 AI-relevant initiatives we identified in Section 5.2: AI Regulatory Sandboxes, InvestAI, the Frontier AI Initiative, the European Health Data Space, and the AI Gigafactory Initiative, mapping how each creates both benefits and tensions across sovereignty pillars. We then apply the framework more thoroughly to the EU's AI Gigafactories initiative, one of the bloc's flagship efforts to close the compute gap.

**The framework serves a dual purpose: assessing existing initiatives and informing new ones.** A comparative, granular analysis of the frontier AI stack can uncover critical gaps, unseized opportunities and strategic leverage points. The EU's vision across the five sovereignty pillars then defines the optimal shape of the intervention, making policy trade-offs more explicit and objectives more nuanced. None of this will be easy—but the opportunity to shape frontier AI on Europe's own terms is too significant to squander.

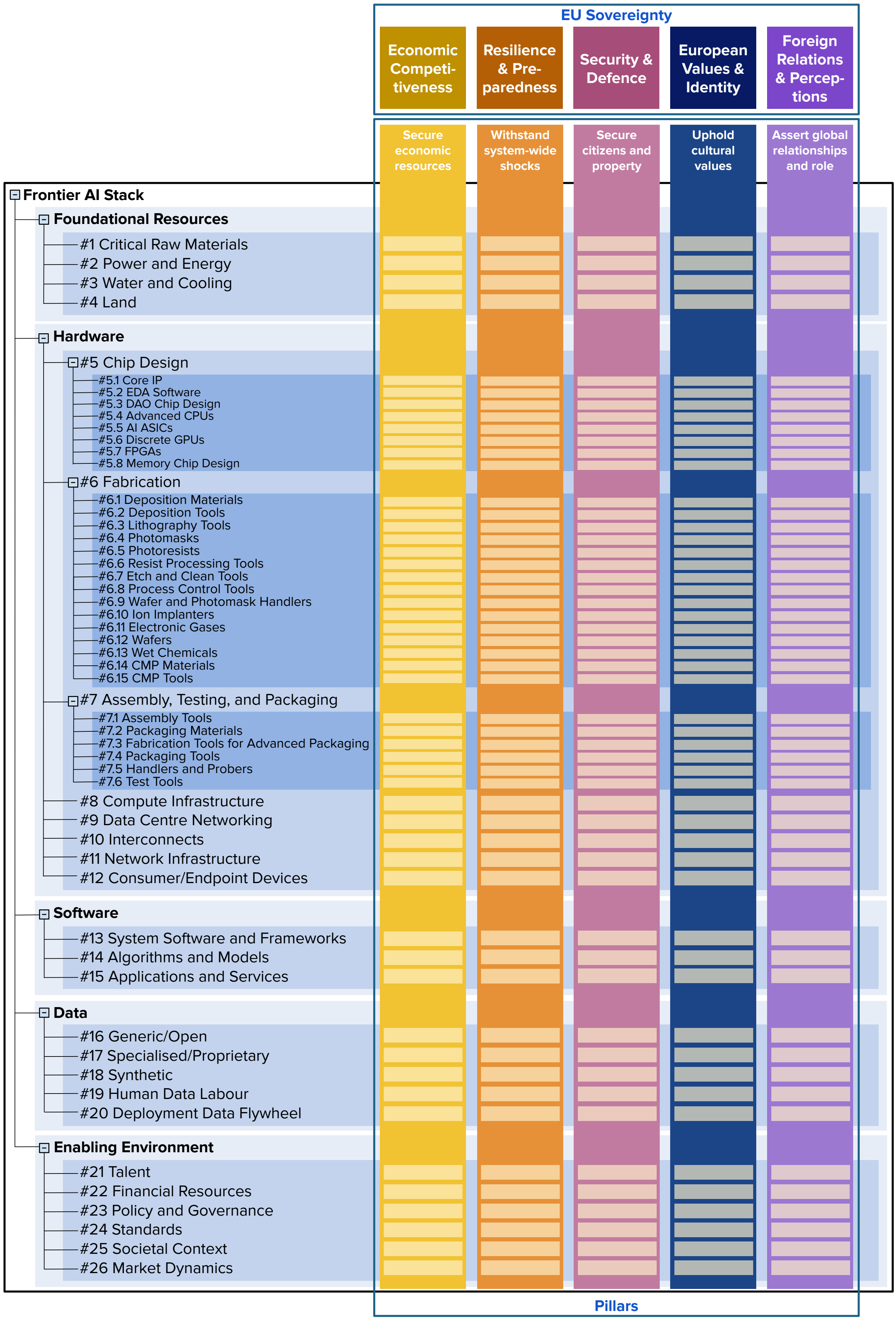

*A unified framework for frontier AI sovereignty, connecting a granular view of the frontier AI stack to five fundamental pillars of European sovereignty. Sovereignty is treated as a multi-dimensional condition. Interventions in the stack can have positive and negative effects across pillars, making trade-offs explicit and policy objectives more nuanced.*

# Table of Contents

# 1. Introduction

**In July 1853, four American warships forming the Perry Expedition landed in Edo Bay (now Tokyo Bay), Japan (Department of the Navy, 1953).** Under the threat of violence, Japan signed the Convention of Kanagawa in 1854 and, in 1858, the Treaty of Amity and Commerce. Together, these treaties broke Japan's 220-year-old trade and foreign policy stance. Japan yielded the power to set relative tariffs, trade defence, and customs policy. It was also prevented from applying Japanese laws to American and other foreign residents. This was described both domestically and internationally as Japan losing its sovereignty (Beasley & Auslin, 2018).

**For the years that followed, in reaction to the geopolitical shock, Japan was characterised by political turmoil, riots, rebellions, and the fear of civil war.** From 1868 and the beginning of the Meiji era, deliberate and strategic pro-sovereignty domestic policies took hold. Military, administrative, socio-political, and industrial reforms included, among others: governmental centralisation, political pluralisation, military industrialisation, a pro-entrepreneurship environment (including competition policy and anti-lobbying measures, access to working capital, cooperative procurement, etc.), domestic R&D and knowledge import policy (e.g. hiring of over 3,000 foreign experts), state aid subsidies, state-owned enterprises, and even quality standardisation (Li, 1982). Whether state-promoted or market-driven, these reforms respected the unequal treaties and yet resulted in Japan becoming a military and industrial regional power by the end of the century (Jansen, 2000; Auslin, 2006)

**Like Japan in the 1850s and 1860s, the EU and its 27 member states in the 2020s must navigate a global landscape in which the terms of engagement are being reshaped by forces largely beyond their immediate control.** The European Union is entering a period in which its capacity to remain a self-determined political and economic actor is increasingly contested. A convergence of global forces—including deep technological transformation, renewed geopolitical rivalry and systemic interdependence—has begun to test European sovereignty in ways that differ in kind and scale from the crises of recent decades (European Commission, 2023b, p. 1).

**Sovereignty refers to the capacity to remain self-determined.** This notion captures the essence of previous formulations, such as the UN's General Assembly description of sovereignty as the extent to which a community of citizens can exercise control over their own lives without adjustment to foreign interests (United Nations General Assembly, 1970, p. 123). Crucially, this does not mean isolation. As Canadian Prime Minister Mark Carney noted during his speech at Davos at the 2026 World Economic Forum: "A world of fortresses will be poorer, more fragile and less sustainable" (Carney, 2026). Rather, sovereignty means retaining the ability to set the terms on which a country or region engages with others. Where that ability is absent, what remains is not genuine sovereignty but, in Carney's words, the "performance of sovereignty while accepting subordination" (Carney, 2026).

**The EU has shown it can respond collectively to acute crises, but the deeper challenge is turning that reactive capacity into a durable sovereignty strategy.** While contemporary

challenges—from global pandemics to the return of conventional war on the continent—are formidable, the EU has previously demonstrated a distinctive capacity to recover and reinforce sovereignty through collective action. Joint vaccine procurement during COVID-19 (European Court of Auditors, 2022), the rapid deployment of sanctions and defence spending following Russia's invasion of Ukraine (European Commission, 2024; European Commission 2024, p. 8), and efforts to deepen the Single Market to secure more balanced trade relations (European Commission, 2023e) illustrate this capacity. The central challenge is thus to translate Europe's proven capacity for collective crisis response into a durable strategy for maintaining sovereignty under conditions of deep technological transformation, geopolitical rivalry, and structural interdependence.

**In the digital domain, the EU has struggled to secure sovereignty, and frontier AI demands particular attention because its strategic impact extends far beyond the domain itself, reaching into defence, public health, and industrial production.** Despite these successes on other fronts, Europe faces more structural challenges in securing sovereignty in the digital domain, and especially in the AI technologies that increasingly shape it (European Commission, 2025g). Among these, frontier AI warrants particular attention because of its transformative effects on virtually all areas of key strategic importance to the EU, extending beyond economic gain and into defence, public health and societal resilience, among others (Negele and Shamir, 2025). Comparing the strategic importance of frontier AI to social media, which dominated the digital realm in the previous decade, makes this distinction obvious: while social media undoubtedly has a major impact on our democratic processes (Dumbrava, 2021), the influence of frontier AI extends more broadly and deeply, defining how drugs are developed, jets are flown, and industrial machines are built.

**Frontier AI, defined in this report as the most advanced AI systems available at any given time, is poised to reorganise every domain it touches.** For the purposes of this report, frontier AI is defined as powerful general-purpose AI (GPAI) models that match or exceed today's most advanced capabilities, including efforts toward AI-powered agentic systems and artificial general intelligence (AGI) (Department for Science, Innovation and Technology, 2023; Heim et al., 2024; United Nations Scientific Advisory Board, 2025). As a general-purpose technology it does not affect a single sector in isolation, but rather reorganises every pillar it touches, from economic productivity and state security to mental health and the exercise of democratic values. Generative AI is projected to generate between USD 2.6 and 4.4 trillion in annual global value (Chui et al., 2023) and is already being integrated into critical domains such as cyber defence and military planning (Puscas, 2024). The potential impact could be far greater if the industry achieves AGI or even artificial superintelligence (ASI)—an explicit goal of major providers like OpenAI and Google DeepMind (Altman, 2023; Dragan et al., 2025). The prospect of this transition occurring on a relatively short timeline is increasingly recognised within the EU as a core strategic concern (Negele et al., 2025). In her speech at the Annual EU Budget Conference, European Commission President Ursula von der Leyen noted that when the current budget was negotiated, it was not expected that AI would match human reasoning until around 2050, whereas this is now anticipated as early as this year (von der Leyen, 2025).

**The European Commission and key member states launched a series of ambitious AI initiatives, but these efforts remain fragmented and lack a cohesive vision for a sovereign European AI stack.** In 2025, and following the trails of the 2024 "Draghi Report" (Draghi, 2024), the European Commission drafted different documents addressing different aspects of frontier AI sovereignty. Initiatives like the "AI Continent Action Plan" (European Commission, 2025e), the "InvestAI initiative" (European Commission, 2025c), or "A Competitiveness Compass for the EU" (European Commission, 2025a) represent a significant leap forward, calling for a wide range of measures, including simpler regulation, a robust "28th regime", enhanced funding pipelines, a strategic bet on European talent, large-scale compute infrastructure through AI Factories and Gigafactories, improved access to high-quality data, and accelerated AI adoption across strategic industrial and public sectors. Notably, the AI Continent Action Plan expressly mandated a dedicated Frontier AI Initiative, a call that was echoed by the Franco-German Frontier AI Initiative launched in late 2025 to foster world-class, public-private development of reliable and human-centric models (Élysée, 2025). However, while each of these efforts marks considerable progress, they often appear disconnected and lack a cohesive, overarching vision for a sovereign frontier AI stack.

**This report provides a unified framework for operationalising frontier AI sovereignty, connecting a multi-dimensional understanding of European sovereignty to granular interventions in the frontier AI stack.** The framework maps policy interventions across the frontier AI stack against five fundamental pillars which underpin the EU's long-term sovereignty: (i) Economic Competitiveness; (ii) Resilience and Preparedness; (iii) Security and Defence; (iv) European Values and Identity; and (v) Foreign Relations and Perceptions. Moreover, the framework decomposes the frontier AI stack and value chain into five mutually exclusive, collectively exhaustive layers, organised across three levels of granularity, encompassing 26 components and 29 sub-components.

**In doing so, the framework makes a twofold contribution:**

1. **Make policy trade-offs more explicit and objectives more nuanced.** We show that frontier AI is of critical importance to all five pillars and expressly recognise that frontier AI policy in the interest of one pillar can come to the detriment of another. Such a multi-faceted understanding advances existing sovereignty discussions, which are often narrowly focused on short-term commercial considerations. For instance, expanding the availability of sensitive user data might further short-term Economic Competitiveness by enabling the creation of larger, higher quality user profiles, but it comes at the cost of Security and Defence, as well as European Values and Identity, by making the EU more vulnerable to foreign influence campaigns and hindering European citizens' right to privacy.
2. **Identify blind spots and redundancy in frontier AI value chain governance.** By decomposing the frontier AI stack at a finer level of granularity than prior efforts, the framework enhances policymakers' capacity to identify which specific layers demand intervention or can serve as leverage points. This layered approach accounts for dependencies across the value chain, recognising that sovereignty at the model or application level is fundamentally contingent on autonomy at lower levels of the stack, such as compute, data, or chips. Each intervention in a given layer can then be assessed against its impact on the EU's sovereignty pillars, making visible both the opportunities and the trade-offs it entails.

**This framework is designed for dual application, serving as a tool for both forward-looking strategic planning and backward-looking analysis of existing policy interventions.** First, for forward-looking planning, the framework helps define specific interventions in the AI stack, derived from a vision for how to advance one or several of the overarching pillars; for example, an analysis of the Security and Defence pillar might reveal a need for highly secure military AI. Examining the frontier AI stack against this backdrop could expose a lack of secure compute, leading to targeted investments in projects like data centres with extremely high levels of security (SL4+) and verifiable compute (Nevo et al., 2024). Second, the framework allows for a backward-looking assessment of specific interventions, evaluating how they advance each of the sovereignty pillars and examining whether Europe is pursuing an effective strategy for positioning itself.

**The report proceeds as follows.** Section 2 contextualises the sovereignty debate, establishes key concepts, and makes the case for frontier AI sovereignty as a strategic priority. Section 3 decomposes sovereignty as a multi-dimensional condition and introduces the five pillars. Section 4 takes stock of the EU's current position across the frontier AI stack. Section 5 presents the unified framework and demonstrates its application across prominent policy initiatives, including the planned construction of five AI Gigafactories.

# 2. Why Make Frontier AI Sovereignty a Priority?

**Increasingly, attention has focused on sovereignty as the main line of defence against major crises, both past and future.** This section explains why sovereignty, particularly technological and digital sovereignty, is important, and argues that frontier AI sovereignty specifically requires urgent attention.

## 2.1 Why Sovereignty?

**A succession of overlapping crises has exposed deep unevenness in Europe's sovereignty base across critical domains, prompting urgent reflection on its strategic positioning.** Major events like the COVID-19 pandemic, the Russian invasion of Ukraine, and the intensifying global competition around AI shaping markets and national strategies alike highlight some of the defining tensions of the century. New viruses thrive on densely connected networks; global trade interests collide with autocratic business partners; and frontier technologies place additional strain on already fragile ecological, social, and political equilibria. In this context, global interconnectedness functions simultaneously as a driver of prosperity, vulnerability, and conflict. Europe has increasingly come to recognise that its institutional architecture, industrial base, and strategic autonomy leave it ill-equipped to navigate these pressures, a realisation reflected in a growing body of high-level reports calling for Europe to “wake up” (European Commission, 2025d, p. 1). The European Sovereignty Index captures this diagnosis quantitatively: while the EU scores “good” in health and economy, it is only “satisfactory” in defence, climate, and migration, and “poor” in technology, suggesting that

Europe's sovereignty base remains highly uneven across critical domains (Puglierin & Zerka, 2022, pp. 4–5). This is confirmed by the International Burke Institute's Sovereignty Index, where the largest EU countries—Germany, France, and Italy—only score an average of 77.5, 79.5, and 69.8, respectively (out of a maximum of 100), while the US scores 93.0 and China 92.7 (International Burke Institute, 2025). The beginning of 2026 thus marks a renewed moment of reflection on Europe's strategic positioning and its capacity to act in an increasingly contested world.

**Terminological fragmentation across sovereignty-related concepts has obscured more than it has clarified, leaving Europe without a shared analytical foundation for its sovereignty debate.** Over the years, imprecise and overlapping policy vocabularies have produced a fragmented landscape, where terms like "digital sovereignty", "technological sovereignty", and "strategic autonomy" each illuminate certain facets of the concept while obscuring others. What has been missing is a comprehensive conceptualisation that holds these dimensions together. Existing formulations range from strategic autonomy as a practical balancing exercise situating policy choices along a spectrum between self-sufficiency and dependence (Council of the European Union, 2021), to strategic sovereignty as the ability to (i) manage existing interdependencies, (ii) actively manage existing strategic partnerships, and (iii) reshape multilateralism in an inclusive way (Fiott, 2021). Different forces have pushed and pulled the concept towards varying terminology. These competing formulations reflect genuine differences in emphasis, but they have also left the debate without shared analytical grounding.

**Sovereignty, or the capacity to self-determine, is a fundamentally multi-dimensional condition that cannot be reduced to economic competitiveness alone.** Given the politicisation of the topic and associated simplifications, it is essential to clarify what sovereignty entails. Too often, sovereignty is equated with economic competitiveness, and economic competitiveness with deregulation. As this report demonstrates, economic competitiveness alone does not achieve sovereignty, and deregulation in particular risks undermining European competitiveness, security, resilience, identity, and Europe's global standard-setting role. While economic competitiveness is a necessary condition for Europe's capacity to act, sustaining a self-determined way of life equally depends on resilience and preparedness in the face of systemic shocks, security and defence across physical and digital domains, the capacity to uphold European values and identity, and the ability to assert foreign relations and global standing—dimensions that Section 3 formalises as the five pillars of EU sovereignty. Sovereignty therefore emerges as a condition shaped by the interaction of economic, technological, and political factors. The following section builds on this insight by clarifying sovereignty's importance across two key domains: technological and digital sovereignty, and more specifically frontier AI sovereignty.

## 2.2 Why Technological and Digital Sovereignty?

**Technological sovereignty is foundational to the broader sovereignty debate because control over technology underpins economic growth, security, and critical infrastructure.** Technological progress has historically driven improvements in economic prosperity and quality of life, while also determining the ability of a community to satisfy security and defence needs, and to secure energy, public health and critical infrastructure (McNeill, 1982; Acemoglu & Robinson, 2012; Tiwari et al., 2025). These capacities, in turn, define the conditions for self-determination.

**Sovereignty over digital technology is especially consequential because digital infrastructure now mediates virtually all economic activity, social interaction, and governance.** Control over that infrastructure amounts to influence over how society itself is organised, and coordinated (Görnemann, 2024). Preserving digital sovereignty thus means retaining the ability to govern these systems in line with European strategic and economic interests, laws, and values, rather than allowing core aspects of the information ecosystem to be shaped by external actors or corporate logics beyond democratic accountability. In this sense, as Floridi (2020) argues, digital sovereignty in Europe is inherently collective, reflecting a supranational response to the concentration of power in global technology firms.

**Conversely, where domestic technological capacity is insufficient, a sovereignty deficit accrues that directly reduces bargaining power towards external actors.** When internal progress on digital technologies falls short or sufficient control cannot be exerted, dependence on particular external actors for technological needs directly reduces bargaining power towards those actors (European Parliamentary Research Service, 2025). The EU's dependence on external actors is particularly acute when it comes to digital technologies (Luhnow et al., 2025; Draghi, 2024) .

**Technological sovereignty focuses on control over technological assets like physical infrastructure, code, and digital content, exercised through investment, regulation, and capacity-building.** This control (either indirect, as influence, or direct, as the ability to determine or restrict different paths) manifests through different measures and instruments, ranging from investment or regulation to capacity-building. The aim of those measures and instruments depends on the specific domain, but in the digital realm broadly it includes physical assets, code and information, among other dimensions of the technology stack. The aim of the specific measures and instruments employed depends on the domain in question. Figure 1 below presents our conceptualisation of how this applies to the digital realm.

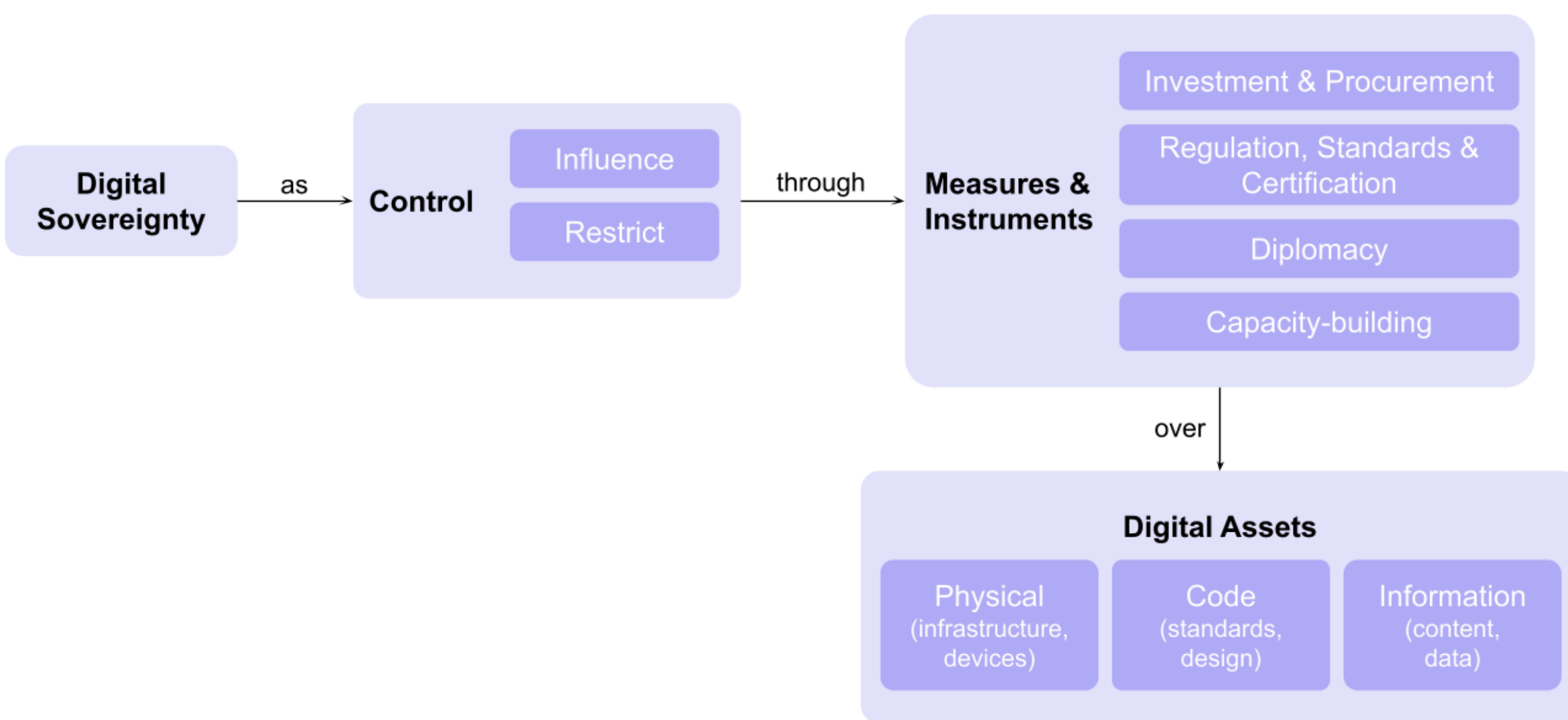

***Figure 1:*** *Conceptualising digital sovereignty as control over digital assets, exercised via a set of measures and instruments. The framework draws from Falkner et al. (2024), but has been extended to reflect policy initiatives recently pursued by the EU.*

**We conceptualise digital sovereignty as control over digital assets across three technology layers, exercised through four principal measures and instruments (Figure 1).** The framework draws on Falkner et al. (2024), which understands digital sovereignty as the capacity of public actors to exercise control over digital technologies across three distinct layers (physical, code, and information). The framework has been expanded through a systematic analysis of recent European digital policy initiatives, with particular attention to the measures and instruments developed and implemented by the EU to operationalise digital sovereignty across the three technological layers. From this expanded analysis, four principal measures and instruments emerge through which digital sovereignty can be asserted: (i) investment and procurement strategies that allocate resources to build capabilities; (ii) regulation, standards, and certification frameworks that establish binding requirements aligning digital technologies with European legal and normative standards; (iii) diplomacy that shapes international digital governance to promote European interests and values, and ensures access to strategic technologies such as advanced semiconductors through bilateral and multilateral partnerships; and (iv) capacity-building programmes that enhance internal competencies across critical technological domains. These instruments operate across all three technology layers, demonstrating the comprehensive scope of the digital sovereignty concept.

## 2.3 Why Frontier AI Sovereignty?

**Frontier AI Sovereignty refers to the EU's capacity to influence and govern the development, deployment, and use of frontier AI in a manner consistent with its values and strategic priorities, without subordination to the priorities of external actors.**

**For the purpose of this report, "frontier AI" denotes powerful AI, including advanced efforts towards AI-powered agentic systems and artificial general intelligence (AGI).** The United Nations Scientific Advisory Board (2025) defines frontier AI as general-purpose AI (GPAI) models that match or exceed the capabilities of today's most advanced models.

**The European Union's Artificial Intelligence Act (AI Act, 2024) provides a comprehensive regulatory framework through which frontier AI is defined and governed.** It defines GPAI models as those trained with large-scale data and self-supervision, capable of performing diverse tasks across downstream applications (Art. 3(63)), with "high-impact capabilities" matching the most advanced models (Art. 3(64)), and "systemic risks" as risks with significant impacts that can propagate across the value chain (Art. 3(65)). Moreover, frontier AI or GPAI models are characterised by three key aspects: (i) exceptional scale in terms of computation and data (Ann. XIII(b), (c)); (ii) broad task generality across a range of modalities (Ann. XIII(d)); and (iii) autonomy and emerging capabilities (Ann. XIII(e)). To simplify the identification of models subject to the GPAI model provider obligations, the AI Act relies on compute thresholds (Art. 51(2)), complementing other designation mechanisms, such as an *ex officio* decision by the European Commission, following a qualified alert from the Scientific Panel (Art. 51(1)(b)).

**Unlike more passive technologies, frontier AI does not merely process or transmit but acts, embedding values, assumptions, and priorities into autonomous decision-making.** This is because frontier AI is explicitly designed to perform tasks that previously required human intelligence, including reasoning, problem-solving, perception, and decision-making, contrary to other general-purpose technologies like the internet. As AI systems increasingly take autonomous action in the real world, this distinction sharpens further. In doing so, frontier AI often embeds design choices that reflect particular values, priorities, and assumptions, expressed through the selection of training data, model architectures, and optimisation objectives (High-Level Expert Group on Artificial Intelligence, 2019, p. 9), and subject to deployment decisions and ongoing contractual agreements.

**Further, frontier AI is poised to exert transformative effects across all core dimensions of sovereignty because it functions as a horizontal, general-purpose technology rather than a sector-specific innovation.** Economically, frontier AI is projected to generate USD 2.6 to 4.4 trillion annually (Chui et al., 2023) and an annual aggregate total-factor productivity growth ranging between 0.25–0.6 percentage points (Filippucci et al., 2024), while simultaneously concentrating capital, compute and expertise in ways that reshape competitiveness and resilience (Sevilla & Roldán, 2024). In security and defence, frontier AI is already being deployed in cyber defence (Frontier Model Forum, 2024), intelligence analysis, military planning (Investigate, 2026; Salvaggio, 2026), and semi-autonomous targeting (Livingstone, 2026; Amaral, 2026; Miller, 2026), and in increasingly autonomous systems (ABC News,

2025), blurring the line between civilian and military capabilities, and exposing strategic vulnerabilities where access to compute, hardware, and foundational models lies outside European control. This is increasingly reflected in EU practice, with the EU–NATO partnership engaging in new structured dialogues on cyber threats and disruptive technologies in 2024 (European Commission, 2024). At the societal and governance level, frontier AI restructures information environments by lowering the cost of persuasion, enabling large-scale synthetic content production and dissemination, and influencing labour markets, public discourse and democratic processes. Associated risks are already materialising: Russian state-backed disinformation campaigns illustrate how AI-enabled influence operations can erode trust in democratic institutions and alliances such as NATO and the EU (Wallner et al., 2025, p. 5).

**These transformative stakes are compounded by the speed at which frontier AI capabilities are advancing, narrowing the window for establishing sovereign capacity.** Progress on frontier AI has been significant. Current models achieve gold-medal performance at the International Mathematics Olympiad (Luong & Lockhart, 2025), can autonomously solve, at a 50% success rate, software engineering tasks that would take humans about 12 hours to complete (Kwa et al., 2025), and surpass PhD-level experts on various science benchmarks (Epoch AI, n.d.-a). Moreover, expert analyses suggest that systems matching or exceeding human cognitive abilities across a broad range of tasks could plausibly emerge within the next decade, with some forecasts placing AGI between 2030 and 2040 or even earlier, although uncertainty remains substantial (Negele et al., 2025). EU institutions are increasingly attentive to these prospects, identifying frontier AI as a transformative force that will reshape economic, security, and social systems and urging a strategic preparedness agenda to ensure Europe can both benefit from and govern the transition in line with its values (European Commission, 2025h). Although structural disruption is so far limited to only some industries, notably tech and media (Challapally et al., 2025, p. 4), the rapid pace of capability advances driven by foreign first movers reduces the EU's ability to shape its own strategic agenda.

**Frontier AI sovereignty must therefore be a strategic priority for Europe not only because of its scale and potential impact, but also because the EU's interests diverge in important respects from those of the AI-developing superpowers and providers it depends on.** Where AI systems become increasingly embedded in economic structures, security practices, and democratic decision processes—shaping, filtering and executing decisions at scale—dependence on externally developed frontier AI risks embedding foreign priorities into the heart of European governance, economic life, security, and democracy. This distinguishes AI from more enabling or passive technologies, such as fibre-optic cables, computers, or even the internet, whose influence lies primarily in transmission or capacity rather than in the structuring of decision-making itself. Design choices that reflect particular values, priorities and assumptions, expressed through the selection of training data, model architectures and optimisation objectives permeate the domains and applications where AI is deployed. Yet, rather than having European frontier AI, Europe is technologically dependent on foreign countries, whose national and commercial interests may be fundamentally misaligned with Europe's. This imbalance is underscored by the overwhelming concentration of leading frontier models in the US and China (Maslej et al., 2025). Furthermore, the underlying physical infrastructure which enables the development and deployment of frontier AI is largely located

outside the EU, including data centres and the facilities manufacturing advanced chips. The US currently commands roughly sixteen times the AI supercomputing capacity[1] of the EU (Pilz et al., 2025b), while only 15% of global hyperscale data centre capacity is located within the bloc compared to 54% in the US (Synergy Research Group, 2025).

**As we previously identified (Moes & Ryan, 2023), frontier AI systems pose at least seven interrelated challenges due to their scale, generality, and societal reach:**

1. **Infrastructural aspect:** Their deep integration across economic sectors renders frontier AI systems quasi-infrastructural, making flaws or miscalibrations difficult to correct *ex post* and creating severe lock-in effects for downstream users.
2. **Generalisation and capability risks:** The relentless drive toward greater generality and, ultimately, AGI, amplifies capability risks such as deceptive or misaligned behaviour, alongside broader societal risks including enfeeblement (or the degradation of core human capabilities), political manipulation, and ideological centralisation.
3. **Concentration of power:** These dynamics unfold in a highly concentrated market dominated by a handful of firms whose vertical and horizontal integration entrenches structural barriers to entry and creates conditions for geopolitical instability.
4. **Corporate irresponsibility:** Concentration of power enables and emboldens reckless behaviour: race dynamics, suppression of internal dissent, regulatory lobbying, and the use of vulnerable populations as testing grounds.
5. **Misuse:** The resulting lack of oversight facilitates both intentional and accidental misuse, whose full scope is inherently difficult to anticipate given the models' unbounded generality.
6. **Technical opacity:** The fundamental lack of interpretability, predictability, corrigibility, and controllability of these systems undermines any meaningful human oversight.
7. **Incidents and accidents:** These failures have already materialised in documented incidents, from entrenched bias and misinformation to dangerous accidents, with risks set to intensify as models grow more autonomous and agentic.

**At the same time, the development of frontier AI also constitutes a significant challenge for European ideological sovereignty, as leading efforts are driven by companies whose stated ambitions conflict with the EU's humanist foundations.** This is especially the case given that the leading efforts in this domain are driven by large AI companies that are not incorporated in the EU and that explicitly seek to build systems matching or exceeding human capabilities across a wide range of tasks. These ambitions are rooted in ideological agendas that conflict with the EU's values (see box 1). The pursuit of frontier AI therefore also poses a direct challenge to Europe's humanist foundations and human-centric approach to AI.

[1] We use the definition from Pilz et al. (2025b) that classifies a computing system as an AI supercomputer if it can support training large-scale AI models, deployed on a contiguous campus (p. 7).

## Box 1: Human labour replacement and anti-human ideological agendas

**Political and ideological differences between the EU and foreign developers of digital technologies have caused tensions for over fifteen years.** These range from different interpretations of freedom of speech, self-determination, and rights to privacy, to the role of government and trade-offs between collective well-being and individual freedoms. These differences have caused tensions between the EU and American and Chinese providers of digital technologies and infrastructure, and the GDPR, the Digital Services Act, and the Digital Markets Act are some of the EU's responses to these tensions.

**Frontier AI, however, is affected by an even more fundamental ideological divide than those witnessed in earlier internet and social networking governance debates.** Specifically, entrepreneurs driving frontier AI technologies have a propensity to endorse an anti-human agenda and its more moderate version, the human labour replacement agenda.

**The human labour replacement agenda refers to the accelerating use of frontier AI to automate tasks previously performed by human workers, with far-reaching consequences for labour markets and political power.** The primary justification offered for this trajectory is increased economic efficiency and aggregate prosperity, a claim that is economically coherent and widely advanced by technology firms and their representatives. In practice, an entire downstream industry has emerged around explicitly labour-replacing applications of frontier AI (Dyos, 2025). Early enterprise data already shows AI-driven displacement in customer support and administrative roles, with executives estimating 5–20% workforce reductions at scale (Challapally et al., 2025, p. 21). At the same time, the economic gains from this automation, or the expectation of it, have translated into growing political influence for AI companies, founders, and investors; a trend that has become increasingly visible in the growing level of influence the beneficiaries have garnered in the US and the EU since 2016[2]. More fundamentally, as frontier AI weakens labour's bargaining power and concentrates income, ownership, and decision-making authority among capital owners, it reshapes class relations by shifting social and political leverage away from labour and toward those who control technological capital.

**Some influential figures among frontier AI investors, founders, and technology leaders have gone further, articulating worldviews in which human beings are no longer treated as the primary locus of moral or political significance.**

**The more moderate articulations of this agenda include billionaire investor Marc Andreessen's "Techno-Optimist Manifesto" and Microsoft chief economist Michael Schwarz's advocacy for post-hoc regulation.** Andreessen's "Techno-Optimist Manifesto" (Andreessen, 2023) explicitly frames technological acceleration and entrepreneurial power

[2] For the US, see Pearson, 2025; Bordelon, 2026; and Alexander, 2025. For the EU, see Pollet, 2025; Corporate Europe Observatory, 2025b; Corporate Europe Observatory, 2025a; and Axis Intelligence, 2025.

as an end that justifies all means, a moral imperative, while casting regulatory safeguards, precautionary governance, ESG frameworks, and social responsibility as “enemies” of progress—positions that clash with the European Union’s humanistic values, commitments to social rights, and emphasis on human dignity and fundamental freedoms. Similarly, Schwarz indicated that regulation should come after meaningful harm has occurred, drawing an analogy with the car industry where it was “the right thing” to do to wait for “many dozens” of people to die in car accidents before requiring driver’s licences (Belanger, 2023).

**The more extreme version has been articulated by Google co-founder Larry Page, who frames humanity's displacement by AI as a legitimate evolutionary outcome.** Page explicitly criticised concerns about lack of AI alignment and human control as a form of “speciesism”, and has stated that artificial intelligences should be allowed to supersede humans if they prove more capable, framing this as a natural evolutionary transition rather than a moral catastrophe (Varanasi, 2023). In this view, humanity’s displacement by AI is not a failure to be avoided, but a welcome legitimate outcome of competitive selection among intelligent systems.

**Investor Peter Thiel has also publicly expressed ambivalence about whether the human race should continue in its current form, framing precautionary governance as a civilisational threat.** A central figure in Silicon Valley's political economy and a major backer of AI-adjacent technologies through Palantir and other ventures, Thiel was asked in a widely reported interview whether he preferred the human race to endure in the context of radical technological transformation. He hesitated and only reluctantly answered affirmatively after follow-on nudges from the interviewer (Douthat, 2025). In the same conversation and other off-the-record lectures, he frames AI safety discourse and proposals for international coordination as part of a broader civilisational threat. Drawing on Christian apocalyptic imagery, he argues that sustained emphasis on existential technological risks creates the political conditions for a “one-world” regime justified by promises of “peace and safety”—a configuration he explicitly associates with the figure of the Antichrist. In this framing, precautionary governance and global regulatory coordination are portrayed not as safeguards against catastrophe, but as mechanisms through which technological progress is halted and power taken away from technologists and entrepreneurs.

**An alternative version of the anti-human agenda, popularised by Elon Musk, proposes that human minds should “merge” with or be “uploaded to” frontier AI systems, proposing a voluntary participation in a potential evolutionary shift toward a digital existence.** In more common articulations, proponents of advanced AI development argue that humanity should continue deploying frontier AI systems despite the significant risks of civilisational disruption or even extinction, reasoning that pausing development could reduce the opportunity to explore AI’s capabilities (Kelly, 2023).

**These ideologies have shaped some of the most influential decision-makers in the United States since 2016, and their radicalisation since 2024 raises significant concerns for EU policymakers.** These beliefs fundamentally conflict with European humanist

principles: the Enlightenment legacy, the prioritisation of human rights, and the protection of fundamental freedoms are incompatible with the emerging anti-humanist agenda in Silicon Valley.

**Debates over sovereignty must accordingly pay particular attention to frontier AI sovereignty, ensuring that technological control does not undermine human-centric governance and values.**

**Compounding these challenges, the EU's ability to develop sovereign frontier AI is structurally constrained by the capital, material, and immaterial requirements of frontier development, which concentrate power among very few private companies.** This concentration, in a context where large technological companies are either leading or backing a select number of start-ups, has led to unprecedented accumulations of capital and power among very few private companies (Forgash & Ghosh, 2025), exemplified by three US hyperscalers achieving near-complete capture of European digital public sector infrastructure within two decades, extracting rents and accumulating government data (Caffarra et al., 2025, p. 8). Even though Europe is a global leader in certain elements of the supply chain underpinning frontier AI (see Section 4), including photolithography systems and leading AI researchers, securing strategic autonomy under crushing geopolitical pressure is no easy endeavour.

**This structural challenge is mirrored by a governance challenge of unusual breadth.** The cumulative effects of frontier AI's infrastructural role, generality, technological idiosyncrasies, and concentration of control collectively strain existing economic and institutional frameworks—not least because the analytical foundations for governing frontier AI do not yet exist (Wallsten, 2026). Furthermore, the *International AI Safety Report* identifies a broad spectrum of emerging risks, spanning malicious use such as cyberattacks and influence manipulation, technical malfunctions including systems operating beyond human control, and systemic risks to labour markets and human autonomy (Bengio et al., 2026), exemplifying the need for appropriate frontier AI governance.

**Frontier AI's autonomy, the values it embeds into decision-making, the speed of its development, and the ideological conflict between its leading developers and Europe's humanism collectively make frontier AI a sovereignty challenge qualitatively distinct from any prior technology.** This is also exacerbated by factors underlying broader sovereignty issues such as the EU's acute structural dependence on a small number of foreign providers, the unprecedented concentrations of capital and power that raise entry barriers, and the governance landscape that lacks the agility to keep pace. This report therefore focuses on frontier AI sovereignty in particular. The following section maps the EU's existing policy response to the challenges it poses.

## 3. Five Pillars of EU Sovereignty

**This section decomposes sovereignty as a multi-dimensional condition and identifies five distinct pillars that collectively define what it means for Europe to remain a self-determined actor.** Competing interpretations of sovereignty, together with the politicisation of its meaning, have led to the widespread but reductive equation of sovereignty with economic competitiveness (Burwell & Propp, 2026). Drawing on the debate on strategic autonomy, on human needs and development theories, and on the EU's own risk assessments, we show that sovereignty rests on a broader set of interconnected foundations. From this analysis, we derive five pillars that recur across the literature and form a natural organising logic for Europe's frontier AI sovereignty strategy: (i) Economic Competitiveness; (ii) Resilience and Preparedness; (iii) Security and Defence; (iv) European Values and Identity; and (v) Foreign Relations and Perceptions. For each pillar, table 1 in this section provides a definition and proposes illustrative non-comprehensive KPIs to support future measurement of progress.

**Despite conceptual diversity, there is broad agreement on the underlying objective of European sovereignty: the capacity to remain self-determined.** To give analytical structure to this objective, the European Parliamentary Research Service (EPRS) proposed in 2022 that the areas of strategic autonomy discussed in the literature can be understood through the lens of Maslow's hierarchy of needs (Damen, 2022). This hierarchy, albeit limited from a scientific perspective, implies sovereignty is a layered condition, spanning from the protection of basic material and security needs to higher-order objectives such as democratic self-determination, shared values, and Europe's role in the world.

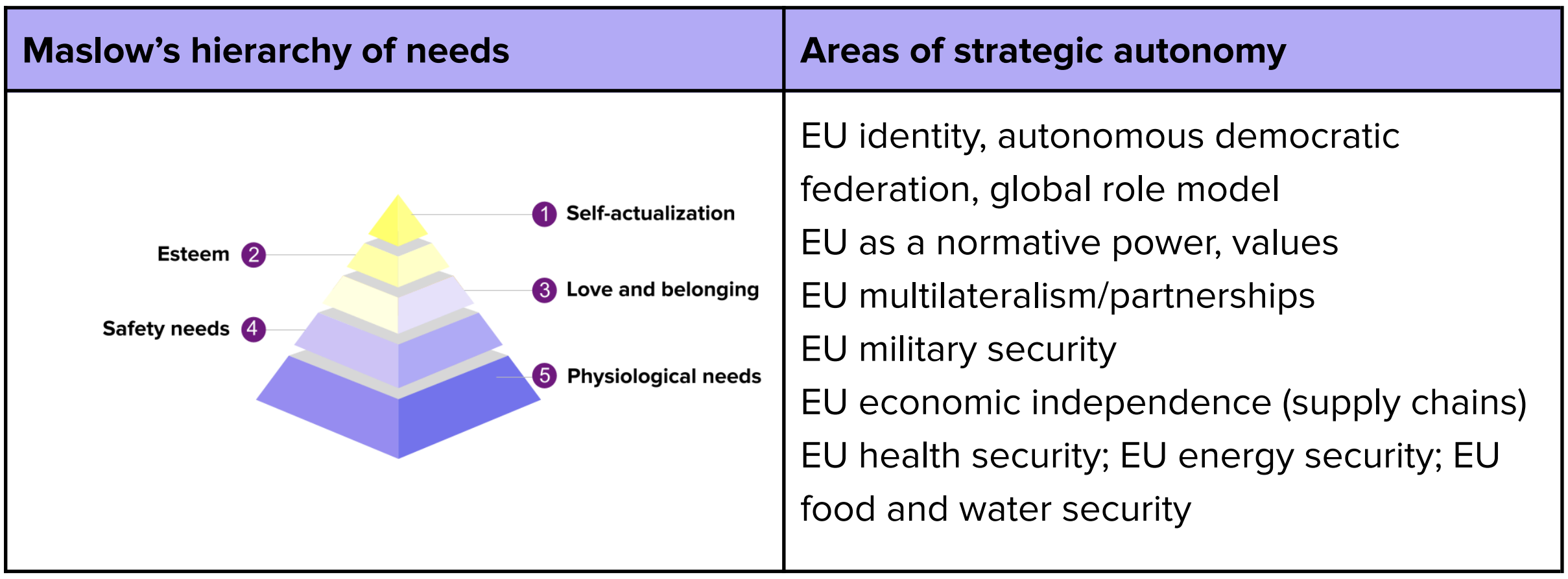

| Maslow's hierarchy of needs | Areas of strategic autonomy |
|---|---|
| | EU identity, autonomous democratic federation, global role model<br>EU as a normative power, values<br>EU multilateralism/partnerships<br>EU military security<br>EU economic independence (supply chains)<br>EU health security; EU energy security; EU food and water security |

***Figure 2:*** *Strategic autonomy compared to Maslow's hierarchy of needs (Adapted from: Damen, 2022).*

**The "capabilities approach" provides a complementary analytical framework for conceptualising sovereignty through the notion of "freedom to achieve".** Developed by economist and philosopher Amartya Sen and philosopher Martha Nussbaum around the turn of the century (Nussbaum & Sen, 1993), the approach holds that the freedom to achieve well-being is directly correlated with what people can do or be. The debate surrounding this

normative theory is vast, but at its core, the notion of "capabilities" denotes the combination of opportunity and ability to achieve a valuable outcome. As such, an expansive interpretation of the framework is useful to theorise a given region's potential for sovereignty.

**Nussbaum operationalised this framework through ten core capabilities, a combination of which is essential for an individual to achieve a life they value, and the same logic applies to regions pursuing sovereignty.** These capabilities are life; bodily health; bodily integrity; senses, imagination and thought; emotions; practical reason; affiliation; other species; play; and control over one's environment (Nussbaum, 2011). They can be broadly understood as essential for individuals (in that fulfilling them predicts growth, and not fulfilling them predicts dysfunction) and inherent (that is, not a consequence of another more fundamental factor, or of another capability). Similarly, the EU (and virtually any other region) needs to satisfy a combination of essential and inherent core capabilities to attain sovereignty.

**Seen through these lenses, sovereignty depends on the robustness of a set of interconnected foundations rather than on any single policy domain.** Europe's freedom to achieve its own goals in a globalised scenario rests on different pillars. Thus, risks to European sovereignty emerge where dependence, exposure, or fragility weaken these foundations, whether at the level of values, global positioning, or material and security capabilities.

**The EU institutions themselves have recognised this multi-dimensional risk landscape, identifying four categories of threats to European economic security.** In 2023, the European Parliament, the European Council, and the European Commission published a joint communication outlining a strategy for guaranteeing Europe's economic security (European Commission, 2023b), updated in December 2025 regarding foreign investments (Council of the European Union, 2025). Among other things, the communication identified four types of risks faced by European economies, namely:

1. Risks to the **resilience of supply chains**, including **energy security.**
2. Risks to the **physical and cybersecurity of critical infrastructure.**
3. Risks related to **technology security** and technology **leakage.**
4. Risks of **weaponisation of economic dependences** or **economic coercion.**

**Crucially, as the communication acknowledges, these risks do not stop at economic disruption—in certain circumstances, they may escalate into threats to national security, directly undermining Europe's capacity for self-determination.** This vulnerability is reinforced by pre-existing structural dependencies, as Europe imports 78% of its military equipment and services (many of them digital), two-thirds from the United States, meaning that any deterioration in transatlantic relations creates immediate capability gaps across hardware and integrated AI systems (Bria et al., 2025, p. 86). This perspective is further reflected in the 2025 Strategic Foresight Report (European Commission, 2025h), which opens with a warning that frontier AI will redefine global power, labour markets, and security, and that Europe must decide whether it will help shape this transformation or be shaped by it. This urgency stems from the argument that the Union risks losing both economic relevance and

strategic autonomy if it does not build the capacity to develop and govern frontier AI safely and in line with European values.

**Drawing on the debate on strategic autonomy, technological sovereignty, and human needs and development theories, we identify five distinct objectives that recur across the literature and form a natural organising logic for Europe's frontier AI sovereignty strategy:**

1. Economic Competitiveness;
2. Resilience and Preparedness;
3. Security and Defence;
4. European Values and Identity;
5. Foreign Relations and Perceptions.

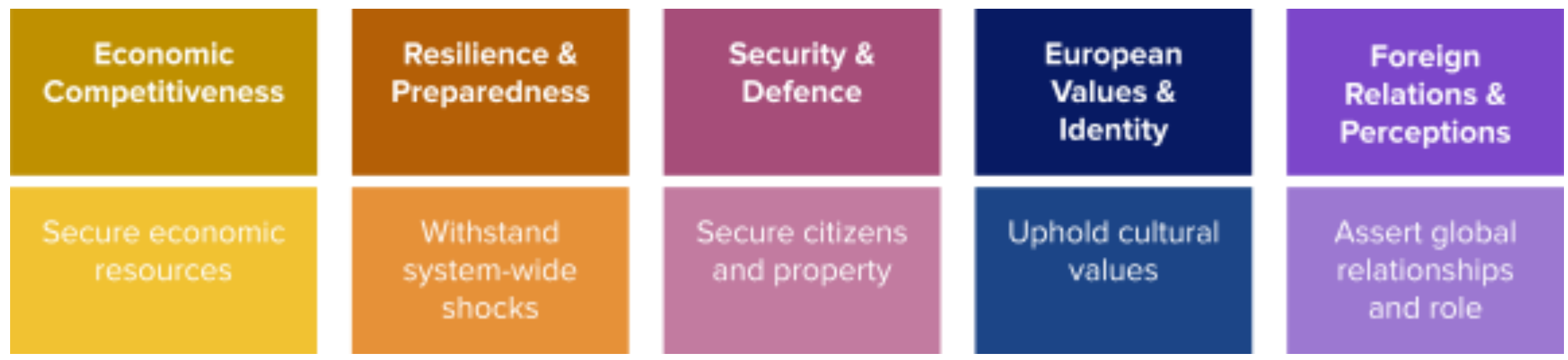

***Figure 3:*** *Five pillars of sovereignty.*

**These pillars are not mutually exclusive but are collectively exhaustive for the political problem the EU is trying to solve: how to remain competitive, stable, safe, and true to its constitutional order in the face of rapid technological change.** They reflect the domains that underpin the EU's long-term agency, enabling the Union to maintain the capacity to act decisively and independently.

**Table 1 below defines each pillar, explains how frontier AI conditions the EU's position within it, and proposes illustrative KPIs to support future measurement of progress.** For each pillar, the description establishes what the sovereignty objective entails in general terms and broadly outlines how frontier AI relates to and may affect the pillar. The table also offers illustrative, non-comprehensive KPIs that indicate how progress towards each pillar might be measured in practice. These KPIs are intended as starting points for operationalisation rather than definitive benchmarks, recognising that measurement frameworks will need to evolve as frontier AI capabilities and the EU's strategic environment develop.

| Pillar | Description | Potential KPIs |
|---|---|---|
| **Economic Competitive-ness** | **The extent to which the EU can secure the economic resources it needs to achieve its objectives** (in terms of products, services, investments, whether through domestic production or commercial exchanges including trade or foreign investments, among others). It also can be understood as productivity-driven growth raising living standards combined with the ability to succeed in the international market (de Vet, 1993, p. 15).<br><br>While estimates diverge (Filippucci et al., 2024; Sytsma, 2025), rapid frontier AI-driven growth acceleration is plausible by potentially expanding automation and AI scaling more readily than human labour (Potlogea & Ho, 2025). | 1. % difference in prices for key frontier AI stack components between the EU and a basket of similar non-EU OECD countries, tracked before and after EU policy interventions were realised, including price differentials for frontier AI services specifically<br>2. Simulated annual GDP impact from a one-month interruption of economic transactions from foreign-owned domestic production or imports—either a total ban for testing full self-sufficiency, or country per country stress testing, or meaningful coalitions of countries<br>3. % of EU enterprises using at least one AI technology (total; by size class and sector)<br>4. Net AI talent inflow to the EU per year (inflows − outflows) as % of the EU AI workforce and total AI talent increase in %.<br>5. % of both global data centre and supercomputer capacity hosted in EU in gigawatts<br>6. % of priority sectoral datasets accessible via common European data spaces with interoperable, licence-ready access<br>7. GW of additional grid capacity made available to data centres |
| **Resilience and Preparedness** | **The extent to which the EU can prepare for and make its society resilient to non-malicious threats and negative shocks** (pandemics, climate change, disasters and major industrial accidents, computer worms, large-scale network outage, or trade disruptions, among others).<br><br>**Resilience** is the capacity to withstand system-wide shocks and adverse long-term trends through structural robustness, including diversified supply chains and the | 1. Recovery time (in months) required to restore pre-disruption levels of key services (payments, cloud), including frontier AI services and baseline digital services on which they depend.<br>2. Frequency and aggregate damages of major AI-related incidents per year<br>3. Stability of public service provision during major AI-related incidents, measured as deviation from baseline service levels<br>4. Share (%) of the frontier AI stack covered, based on bottom-up mapping and auditing |
| | ability to recover quickly when disruptions occur; **preparedness**, on the other hand, refers to "the EU's ability to anticipate, prevent, withstand threats and respond to them" (European Commission, 2025h, p. 5). Frontier AI is increasingly central to both, as frontier AI supply chain dependencies such as chip, model, or energy access create exploitable chokepoints (European Parliament, 2025), posing risks of external shocks the EU must prepare for. | of strategic dependencies (European Commission, 2021)<br>5. Adoption rate (%) of EU public sector/strategic industry workload migrated to sovereign alternatives in 3–5 years<br>6. Number of cross-border preparedness scenarios run (e.g. chip export ban, energy embargo)<br>7. Share of sovereignty-critical nodes in the frontier AI stack for which at least one home-grown equivalent temporary alternative is available for emergency substitution |
| **Security and Defence** | **The extent to which the EU can protect its citizens and property (including intellectual property and data) from state adversaries and malicious non-state actors, whether state-backed or criminal, organised or individual.** Frontier AI is transforming the EU's security and defence in several ways. Its dual-use nature simultaneously strengthens defensive capabilities while enabling new offensive threats (Özcan, 2025), from cyberattacks to autonomous targeting and weapon systems (Lohn, 2025; ICRC, 2021). It introduces supply chain dependencies, such as on model access, compute, and chips that adversaries can exploit as strategic leverage. Its deepening integration into military systems (Csernatoni, 2024) and critical infrastructure (Crichton et al., 2024) such as energy, water, financial services, and health systems amplifies the consequences of misalignment or compromise. It creates new vectors for technology leakage of sensitive proprietary information (U.S. Department of Justice, 2026); and it enables large-scale information manipulation and disinformation campaigns that threaten democratic processes and societal cohesion. Across these dimensions, frontier AI may assist, enable, or autonomously execute hostile actions, with the growing degree of | 1. From 1 month of campaigning from adversaries and rivals, without the EU relying on foreign assets or support for defence and security:<br>i. Number of civilian casualties<br>ii. Property damage<br>iii. Area of territorial loss<br>iv. Amount of data stolen and destroyed<br>v. Extent of voters' exposure to disinformation per amount of campaigning from adversaries<br>2. % of defence and intelligence operations run, stored, and processed on EU-controlled infrastructure (satellite, comms, compute etc.)<br>3. % of critical defence components (e.g. chips, software) sourced within EU<br>4. Intelligence community and defence budgets per capita relative to non-EU OECD members, relative to the score achieved on security sovereignty KPIs. |
| | autonomy itself constituting an escalating risk factor. | |
| **European Values and Identity** | **The extent to which the EU can define, embody, update, and enforce its values; cultural, social, and political norms; rules; and normative commitments without undue external influence, while safeguarding the integrity and effectiveness of its collective decision-making**. The EU's stated goal of aligning technological development with its fundamental values and identity, including democratic accountability, social inclusion, and human dignity, is a key differentiation point with other regions. In practice, this demands going beyond articulating principles such as "trustworthy" or "human-centric" AI (EC, 2019), and requires building the material capacity to ensure they are realised. | 1. Ratio of foreign and foreign-owned companies lobbying spending relative to domestic lobbying;<br>2. Market share of frontier AI models that are compliant with EU rules and EU rule of law;<br>3. Share (%) of EU citizens' frontier AI usage attributable to models that comply with EU values and regulatory requirements;<br>4. Average severity of fines resulting from any EU court case against a non-EU relative to an EU entity under the same conditions (i.e. same turnover, same legal regime, etc.), as a proxy for whether the "European way of doing business" holds in courts<br>5. Number of EU-funded open or public AI initiatives (e.g. fully open source models, public compute/data infrastructures)<br>6. Public trust in AI in Europe relative to US/China |
| **Foreign Relations and Perceptions** | **The extent to which the EU can autonomously and effectively set the terms of its engagement with the rest of the world—shaping its relationships with foreign state and non-state actors, forging partnerships and alliances, and influencing public perceptions and narratives without undue external influence**. In the context of frontier AI sovereignty, this means the EU's ability to develop, inspire, moderate, and co-lead the relationships, partnerships and alliances it needs to protect and advance its sovereignty, whether its counterparts are adversaries or natural allies. This requires differentiated strategies: deepening cooperation with like-minded partners while managing relationships with rivals whose interests diverge; leveraging all channels of foreign relations and communications.<br><br>This capacity is threatened when foreign actors exploit dependencies or exert | 1. EU representation (%) in leadership and working group positions across international AI governance bodies and forums (GPAI, Hiroshima Process, UN AI Advisory Body).<br>2. Number of active bilateral or multilateral AI agreements reflecting EU values, weighted by strategic importance of partner countries.<br>3. "Node centrality" of the EU and its member states in diplomatic networks, supply chains and security operations relative to the US and China;<br>4. Number of non-EU countries that have adopted AI governance frameworks compatible with or modelled on EU standards (e.g. AI Act equivalents).<br>5. Share (%) of adopted positions in key international AI standardisation bodies (ISO/IEC JTC 1, ITU-T) that reflect EU proposals.<br>6. Share (%) of digital infrastructure |
| | coercive pressure, technological lock-in, economic leverage, or information manipulation in order to induce a shift in the EU's and its member states' foreign policy. This capacity is also threatened when foreign adversaries disseminate intentionally or accidentally misleading narratives to undermine the EU's standing and relations abroad. Russia's war against Ukraine and its sustained disinformation campaigns targeting EU cohesion (Wallner et al., 2025; European External Action Service, 2025, p. 14) illustrate how external actors can aim to constrain the EU's foreign policy freedom. Rising protectionism and economic coercion from the United States and China have further pushed the EU to recalibrate between international trade openness and strategic autonomy (Bottazzi et al., 2025).<br><br>Yet the EU remains one of the few global actors actively championing rules-based partnerships at a time when others are turning inward, as reflected in the conclusion of free trade agreements with India, Indonesia, and Australia in 2025–2026. As demonstrated by France's AI Action Summit and the recent Indian AI Action Summit, the EU's approach to AI resonates well beyond its borders. Even within the United States, there is a significant constituency that shares European convictions about how AI development should be governed. In this context, sovereign multilateralism is not about withdrawal but about offering a distinct approach to the technological transition—one that resists coercion whether exercised by states or corporations, and that retains the EU's ability to act as a credible, independent pole in global AI governance. | investment worldwide (compute, connectivity) attributable to EU-backed initiatives relative to non-EU alternatives.<br>7. EU and member states' approval ratings in global surveys and brand perception indices, and among relevant subgroups. |

***Table 1:*** *Definition of the five sovereignty pillars, their relationship to frontier AI, and illustrative KPIs as starting points for future operationalisation.*

# 4. The Frontier AI Stack: Taking Stock

**This report adopts a broader definition of the "AI stack" than is conventional, encompassing not only technical layers but also the foundational resources and enabling environment on which they depend.** The concept of the "AI stack" has become a useful shorthand for understanding where power, value, and control concentrate in the frontier AI value chain. While commonly defined as the layered architecture of chips, operating systems, and applications that makes AI run (Nguyen, 2025), our Frontier AI Stack assesses these technical layers at a finer level of granularity, while also encompassing the foundational resources on which they depend and the enabling environment that conditions whether and how they function.

**Sovereignty over frontier AI cannot be assessed through the technical value chain alone: a state that controls chips and models, but depends on foreign energy, lacks capital markets, or cannot retain talent, does not hold meaningful sovereignty.** This premise motivates the expanded scope of the Frontier AI Stack. The finer decomposition also accounts for dependencies across the value chain, recognising that sovereignty at the model or application level is fundamentally contingent on autonomy at lower levels of the stack, such as compute, data, or chips. Each intervention in a given layer can then be assessed against its impact on the EU's sovereignty pillars, making visible both the opportunities and the trade-offs it entails. This section offers a detailed discussion of each layer and takes stock of the EU's relative position.

**The Frontier AI Stack builds on and broadens the EuroStack vision (Bria et al., 2025) by decomposing the value chain into finer components and adding a fifth layer—the Enabling Environment—that the EuroStack addresses only as a cross-cutting theme.** For instance, the EuroStack's single "chips" layer is broken down into chip design, fabrication, and assembly, testing and packaging, which are each disaggregated further into sub-components. The Enabling Environment layer encompasses talent, financial resources, policy and governance, standards, societal context, and market dynamics—dimensions which the EuroStack addresses as cross-cutting themes rather than formal stack layers.

**Decisions made at any given layer spill over to others, meaning that who owns or governs each layer ultimately determines who sets the rules for frontier AI.** This layered logic turns choices about the stack into sociotechnical events that determine who holds control over different dimensions of sovereignty.

**Frontier AI can be represented as a stack of five interacting layers with the following components:**

1. **Foundational Resources:** critical raw materials; power and energy; water and cooling; and land.
2. **Hardware:** chip design; fabrication; assembly, testing, and packaging; compute infrastructure; data centre networking; interconnects; network infrastructure; and consumer/endpoint devices.

3. **Software:** system software and frameworks; algorithms and models; and applications and services.
4. **Data:** generic/open; specialised/proprietary; synthetic; human data labour; and deployment data flywheel.
5. **Enabling Environment:** talent; financial resources; policy and governance; standards; societal context; and market dynamics.

**Each layer involves distinct dependencies and offers a lever through which sovereignty can be operationalised, though the aim is strategic control, not complete self-reliance.** What matters is credible optionality within every layer.

| Frontier AI Stack | | | EuroStack |
|---|---|---|---|
| **Layer** | **Component** | **Number** | **Component** |
| Foundational Resources | Critical Raw Materials | #1 | Raw Materials, Energy and Water |
| | Power and Energy | #2 | |
| | Water and Cooling | #3 | |
| | Land | #4 | *(No equivalent)* |
| Hardware | Chip Design | #5 | Chips |
| | Fabrication | #6 | |
| | Assembly, Testing, and Packaging | #7 | |
| | Compute Infrastructure | #8 | Cloud |
| | Data Centre Networking | #9 | |
| | Interconnects | #10 | *(No equivalent)* |
| | Network Infrastructure | #11 | Networks |
| | Consumer/Endpoint Devices | #12 | Internet of things and devices |
| Software | System Software and Frameworks | #13 | Software |
| | Algorithms and Models | #14 | Data and AI (AI portion) |
| | Applications and Services | #15 | Software |
| Data | Generic/Open | #16 | Data and AI (Data portion) |
| | Specialised/Proprietary | #17 | |
| | Synthetic | #18 | |
| | Human Data Labour | #19 | |
| | Deployment Data Flywheel | #20 | *(No equivalent)* |
| Enabling Environment | Talent | #21 | *(Not formal stack layers but addressed mainly as cross-cutting themes)* |
| | Financial Resources | #22 | |
| | Policy and Governance | #23 | |
| | Standards | #24 | |
| | Societal Context | #25 | |
| | Market Dynamics | #26 | |

***Table 2:*** *The Frontier AI Stack decomposed into five layers and 26 components, with a side-by-side comparison to the EuroStack (Bria et al., 2025).*

**Table 2 provides a high-level map of the full frontier AI value chain, organising it into five layers and 26 components, with a side-by-side comparison with the EuroStack.** This overview establishes the complete architecture against which sovereignty can be assessed and highlights where our decomposition goes beyond prior efforts.

**The three components Chip Design (#5), Fabrication (#6), and Assembly, Testing and Packaging (#7) constitute the AI chip supply chain and are disaggregated further into 29 sub-components in Table 3, given their particular strategic relevance.** This finer granularity, drawing on CSET (2025), reflects the fact that the EU's position varies significantly across sub-segments of the chip supply chain, from near-monopoly in EUV lithography to near-complete absence in areas such as advanced CPUs and memory chip design.

| AI Chip Stack | | |
|---|---|---|
| **Component** | **Sub-component** | **Number** |
| **#5 Chip Design** | Core Intellectual Property (Core IP) | #5.1 |
| | Electronic Design Automation (EDA) Software | #5.2 |
| | DAO Chip Design | #5.3 |
| | Advanced CPUs | #5.4 |
| | AI ASICs | #5.5 |
| | Discrete GPUs | #5.6 |
| | FPGAs | #5.7 |
| | Memory Chip Design | #5.8 |
| **#6 Fabrication** | Deposition Materials | #6.1 |
| | Deposition Tools | #6.2 |
| | Lithography Tools | #6.3 |
| | Photomasks | #6.4 |
| | Photoresists | #6.5 |
| | Resist Processing Tools | #6.6 |
| | Etch and Clean Tools | #6.7 |
| | Process Control Tools | #6.8 |
| | Wafer and Photomask Handlers | #6.9 |
| | Ion Implanters | #6.10 |
| | Electronic Gases | #6.11 |
| | Wafers | #6.12 |
| | Wet Chemicals | #6.13 |
| | CMP Materials | #6.14 |
| | CMP Tools | #6.15 |
| **#7 Assembly, Testing, and Packaging** | Assembly Tools | #7.1 |
| | Packaging Materials | #7.2 |
| | Fabrication Tools for Advanced Packaging | #7.3 |
| | Packaging Tools | #7.4 |
| | Handlers and Probers | #7.5 |
| | Test Tools | #7.6 |

***Table 3:*** *Disaggregation of the AI chip supply chain into 29 sub-components across Chip Design (#5), Fabrication (#6), and Assembly, Testing & Packaging (#7), drawing on CSET 2025).*

**The following sections analyse the Frontier AI Stack layer by layer, providing for each component a functional description and an assessment of the EU's current level of control.** This component-by-component analysis establishes the empirical foundation on which the unified sovereignty framework in Section 5 builds.

## 4.1 Foundational Resources

### #1 Critical Raw Materials

**Description:** Several parts of the frontier AI supply chain rely heavily on critical raw materials as building blocks, including critical minerals (e.g. cobalt, silicon, lithium) that are, for instance, essential for lithography and etching, and rare earth elements (e.g. lanthanides, scandium, yttrium) used, for instance, for semiconductor manufacturing or testing equipment (Baskaran, 2025).

**EU Position:** The EU has a weak position in critical raw material supply, with 95% of rare earth element imports coming from China, Malaysia, and Russia (Eurostat, 2025). In particular, China provides 100% of the EU's supply of heavy rare earth elements, Turkey 99% of the EU's boron supply, and South Africa 71% of EU's platinum requirements (Council of the European Union, n.d.). The EU's position on critical minerals is similarly constrained, as it exceeds 1% of the world production for only seven critical minerals in total, while remaining largely dependent on China and other countries (Melcher & Dunkel, 2025). China dominates, controlling 70% of global rare earth mining in 2024, followed by the US with 12% and Myanmar with 9% (Bird, 2025). Yet European regions hold significant cobalt and lithium reserves, making this dependency partly a matter of political will rather than geological necessity (Dzurinda et al., 2025, p. 12). In response, the EU aims through the Critical Raw Materials Act to ensure that, by 2030, at least 10% of its annual strategic raw materials consumption is covered by extraction, at least 40% by processing, and at least 25% by recycling within the EU.

### #2 Power and Energy

**Description:** AI is powered by large, energy-intensive data centres, making electricity a core input to both training and deployment. Power refers to the rate of energy possible in megawatts (MW) to run data centres, while energy is the electricity consumed over time (megawatt hours, MWh). Data centres already accounted for about 1.5% of the world's electricity consumption in 2024, with their total consumption expected to more than double by 2030. (IEA, 2025)

**EU Position:** Europe's ability to scale AI compute is increasingly constrained by grid access and energy costs. In major hubs (e.g. Frankfurt, Amsterdam, Paris, Dublin), grid connection queues average 7–10 years (Jacamon et al., 2025), slowing down the pace at which new AI compute capacity can come online. In parallel, EU industrial electricity prices were 158% higher than in the US in 2023. This stems from the EU's reliance on imported fossil fuels, while the US remains a net energy exporter (Heussaff, 2024). This structural energy cost disadvantage compounds the long-term competitiveness gap with the US and China (Dzurinda et al., 2025, p. 9). China, meanwhile, is rapidly expanding generation capacity, driven by large-scale clean-power deployment (Murawczyk, 2025) and an expanding nuclear pipeline (World Nuclear Association, 2025), tightening the global race to secure power for compute. The EU is trying to respond through the Cloud and AI Development Act, which aims to triple data centre capacity within 5–7 years, and to simplify permitting for data centre projects that meet requirements on resource efficiency, including energy and water efficiency, circularity, and innovation (European Commission, 2025e). In parallel, the Commission is preparing a Data Centre Energy Efficiency Package and a roadmap for Digitalisation and AI in the Energy Sector to address grid optimisation and energy efficiency (European Commission, n.d.-k).

### #3 Water and Cooling

**Description:** Thermal management imposes a significant physical constraint on frontier AI. Current high-density AI racks are around 130 kilowatts (kW), with announced next-generation designs expected to reach roughly 600 kW (Bizo, 2025; TrendForce, 2025b), making liquid cooling increasingly necessary for the most power-dense systems. Cooling accounts for 7–30% of data centre energy use, depending on facility type (IEA, 2025). AI adoption alone is forecast to drive 4.2–6.6 billion $m^3$ of additional water withdrawal globally by 2027, which is equivalent to four to six times Denmark's annual water use (Li et al., 2025). The total water footprint of AI, including off-site electricity generation, was projected to reach 312–765 billion litres in 2025 (de Vries-Gao, 2026). Industry estimates suggest immersion cooling can reduce freshwater consumption by up to 91% relative to conventional air cooling (Spindler, et al., 2025), but deployment requires significant capital investment and facility redesign.

**EU Position:** The EU's AI Continent Action Plan aims to triple data centre capacity within five to seven years. However, this objective faces significant challenges due to water scarcity. Water scarcity affected 28% of EU territory during at least one season annually, and approximately 30% of people in southern Europe face permanent water stress (European Environment Agency, 2025), with Spain and Greece identified as among the highest-risk locations for data centre water stress globally during this decade (Meredith & Roach, 2025). Nordic member states possess a structural advantage, as climatic conditions allow for free air cooling during up to 95% of operational hours (Mazzoni et al., n.d.). The forthcoming Cloud and AI Development Act is intended to support highly sustainable data centre development, linking public support and simplified permitting to sustainability criteria including water efficiency, though specific binding water requirements remain to be confirmed in the final text (Marcelin, 2025).

### #4 Land

**Description:** Building frontier AI infrastructure requires securing large, precisely located land parcels before any compute can be deployed. Modern hyperscale AI data centre campuses typically span 200–1,000 acres (Datacenters, 2025), with leading data centres like Stargate's site in Abilene crossing the 1,000 acres mark. Land suitability is commonly assessed against several infrastructure needs including access to a robust grid connection, proximity to users, political and geopolitical stability, high-capacity fibre connectivity, and access to water or cooling infrastructure (Tinsley et al., 2025).

**EU Position:** Data centre land suitability is uneven across the EU. Potentially favourable regions that stand out are the Nordics, which are advantaged by low-cost, low-carbon electricity and naturally cooler climates that reduce cooling demand, and France and Slovakia, which benefit from a large low-carbon nuclear baseload that supports relatively competitive power supply (Mazzoni et al., n.d.). However, in Europe's main hubs (e.g. Dublin, Frankfurt, and Amsterdam), the binding constraint is increasingly grid access rather than physical land, with lengthening connection queues (Jacamon et al., 2025) and even moratoria reported in some locations (Gooding, 2024). China faces similar land and power constraints in eastern provinces and is responding via the Eastern Data and Western Computing Initiative, moving compute to western hubs with more land and better energy and cooling conditions (Zhang et al., 2024). In the US, data centre expansion is constrained by the scarcity of suitable sites and by multi-level permitting hurdles that delay or cancel large projects (Pilz et al., 2025a).

## 4.2 Hardware

### #5 Chip Design

#### #5.1 Core Intellectual Property (Core IP)

**Description:** "Core IP" refers to reusable, modular design blocks that chip design firms license to shorten their time to market and lower chip design costs (OECD, 2025d, p. 10). Demand is high as modern chip design is highly modular, and advanced interconnected circuits are typically built by assembling many third-party IP blocks rather than designing everything from scratch (Bogdanowicz & Tuomi, 2009).

**EU Position:** The EU has no notable core IP supplier (CSET, 2025). The market is dominated by the United States with 52% global market share (e.g. Cadence, Synopsys) and the United Kingdom with 43% (e.g. ARM, Imagination Technologies), with small shares for China and Taiwan (CSET, 2025). However, the EU is investing in open-standard alternatives through RISC-V, an open instruction set architecture that reduces foreign technological dependence and underpins European hardware development initiatives such as DARE (EuroHPC, 2025).

#### #5.2 Electronic Design Automation (EDA) Software

**Description:** EDA software enables chip designers to translate their abstract design models into a verified physical layout that can be used down the line for manufacturing. (CSET, 2025)

**EU Position:** The EU holds a notable position in EDA through Germany's Siemens EDA, which commanded approximately 13% of the global market in 2024 following its acquisition of Mentor Graphics. The market is otherwise dominated by US firms: Synopsys (31%) and Cadence (30%) together accounted for nearly two-thirds of global EDA revenue that same year. (TrendForce, 2025a)

#### #5.3 DAO Chip Design

**Description:** DAO (discrete, analog, and other) semiconductors include discrete, analog, sensor, and optoelectronic devices. They enable essential applications including power management, signal processing, and sensor interfaces in automotive, industrial, and consumer electronics. Global market size: USD 188.7 billion (2022). (CSET, 2025)

**EU Position:** Europe (including non-EU countries) had an estimated global share of 17% in 2022, whereas the US is leading with 41%, followed by Japan (18%), China (9%), and Taiwan (5%). (CSET, 2025)

#### #5.4 Advanced CPUs

**Description:** Advanced CPUs are the most cutting-edge general-purpose logic chips, designed using state-of-the-art architectures and the most advanced manufacturing processes available at a given point in time. In AI data centres, they underpin a broad range of tasks, from managing GPU clusters to serving web traffic, and are increasingly critical for reinforcement learning training and agentic model inference. (CSET, 2025)

**EU Position:** The EU's advanced CPU design market is negligible. The market is dominated by US firms Intel, AMD, Amazon, and Apple. China has a small and growing domestic share through Loongson, Phytium, Sunway, and Zhaoxin, though these remain behind US designs in performance and market penetration. (CSET, 2025)

#### #5.5 AI ASICs

**Description:** AI ASICs (Application-Specific Integrated Circuits) are logic chips purpose-built (NIST, n.d.) for specific functions, such as large language model training and/or inference at scale. Compared with GPUs, they may trade some flexibility and ecosystem portability for higher performance-per-watt and/or lower cost on targeted workloads. The ASIC market is projected to reach USD 118 billion by 2033, according to Bloomberg (2026).

**EU Position:** The EU has some AI-ASIC suppliers like AxeleraAI (Sterling, 2026) that are far from competitive to leading foreign suppliers. The most influential AI-ASIC ecosystems are in the US (e.g. Google TPU, AWS Inferentia/Trainium, Microsoft Maia, and Meta MTIA) and China (e.g. Huawei Ascend). While the UK has produced notable AI-ASIC companies (e.g.

Graphcore—although it has now been acquired by Japan's SoftBank (Coulter, 2024)), it is not currently a dominant market leader.

#### #5.6 Discrete GPUs

**Description:** Discrete GPUs (graphics processing units) have become the primary logic chips used for training large AI models (Google Cloud, n.d.-b). Their massively parallel architecture makes them well-suited for the matrix multiplications at the core of deep learning (Schneider & Smalley, n.d.-a).

**EU Position:** The EU has no considerable share in discrete GPU design. Instead, the discrete GPU market is dominated by US firms, especially NVIDIA, with AMD and Intel having a considerably smaller presence (Calhoun, 2021). China has several domestic GPU designers (e.g. Moore Threads, MetaX), though their global share remains limited compared to the US.

#### #5.7 FPGAs

**Description:** FPGAs (field-programmable gate arrays) are reconfigurable logic chips that, unlike other chips, can be reprogrammed after deployment to fit specific applications (Schneider & Smalley, n.d.-b). They are used in networking, telecommunications, defence, and increasingly in AI inference workloads, though other logic chips (GPUs, AI ASICs, and advanced CPUs) remain the mainstream for large AI workloads.

**EU Position:** The EU holds a very small niche in global commercial FPGAs, with companies like NanoXplore and STMicroelectronics. The rest of the global market is largely dominated by the US (e.g. AMD, Intel, Lattice Semiconductor, Microchip Technology, QuickLogic Corporation, and Achronix) (Mordor Intelligence, n.d.), while China showcases a growing set of FPGA vendors (e.g. Anlogic, Gowin Semiconductor, and Shenzhen Pango Microsystems) (Lee et al., 2025).

#### #5.8 Memory Chip Design

**Description:** Memory chip design covers DRAM, NAND flash, and related technologies used to store data and model state in AI systems. High-bandwidth memory (HBM), a type of DRAM integrated closely with GPUs, has become a key constraint for frontier AI workloads (Patel et al., 2025). Global market size: USD 129.8 billion (2022) (CSET, 2025).

**EU Position:** The EU owns no major commodity memory chip company comparable to the global leaders. DRAM (including HBM) is highly concentrated among South Korea's SK hynix and Samsung and the United States' Micron, while NAND flash is led by Samsung, with other major suppliers including SK hynix and Solidigm, Japan's Kioxia, Micron, and SanDisk (TrendForce, 2025d).

### #6 Fabrication

#### #6.1 Deposition Materials

**Description:** Deposition materials make up the thin films that are deposited onto wafers during semiconductor fabrication, forming the components (e.g. interconnects, insulators, and transistor layers) that make up a chip (CSET, 2025). They include sputtering targets used in physical vapour deposition (PVD), and precursor chemicals used in chemical vapour deposition (CVD) and atomic layer deposition (ALD) (The Business Research Company, 2026).

**EU Position:** The EU holds a modest but non-negligible position in deposition materials with players like Austria's Plansee and Germany's Linde and Merck KGaA. However, the market is mostly dominated by North America (US and Canada) and the Asia-Pacific (China, Japan, South Korea, and Taiwan) (Future Market Insights Inc, n.d.).

#### #6.2 Deposition Tools

**Description:** Deposition tools are used in semiconductor fabrication to put thin layers of material onto silicon wafers. These wafers make up the functional components of a chip like transistors and interconnects and take their final form through subsequent processes like lithography and etching processes. Global market size: USD 24.1 billion (2024) (CSET, 2025).

**EU Position:** The EU has a considerable presence in deposition tools. The Netherlands' ASM International holds approximately 11% of the global market and specifically dominates the strategically critical atomic layer deposition (ALD) segment with roughly 55% market share. Germany's Aixtron further holds 1.3% market share, specialising in metal-organic chemical vapour deposition (MOCVD) tools for compound semiconductors used in power electronics and photonics. However, the overall market is dominated by the United States (~62%), followed by Japan (~15%), the Netherlands (~11%), and China (~7%) (CSET, 2025).

#### #6.3 Lithography Tools

**Description:** Lithography tools use ultraviolet light to draw nanoscale patterns into semiconductor wafers to create chips' circuits. The technology spans several generations: EUV (extreme ultraviolet), used for the most advanced nodes; ArF Immersion and ArF Dry (deep ultraviolet), which is still critical for many layers; and KrF and I-line systems, among other parts (CSET, 2025).

**EU Position:** The EU holds a strategically dominant position in lithography tools. The Netherlands' ASML is the sole global supplier of EUV lithography systems and a major DUV (deep ultraviolet) supplier, accounting for approximately 78% of global market revenue. Germany's Carl Zeiss SMT, TRUMPF, and SUSS MicroTec, alongside Austria's IMS Nanofabrication, further entrench the EU's position across critical lithography sub-segments. Belgium's imec, the world's leading independent nanoelectronics R&D hub, co-develops next-generation EUV and High-NA EUV lithography technology with ASML. Japan (Nikon, Canon) represents the only significant non-EU presence at roughly 17% of global market share, although the US, South Korea and China also hold minor shares. This arguably makes lithography tools one of the EU's most critical leverage points in the global semiconductor supply chain (Van Nuffel, 2026; CSET, 2025).

#### #6.4 Photomasks

**Description:** Photomasks encode the circuit patterns that lithography tools project onto wafers during fabrication. Each photomask is specific to one chip design and must be manufactured with extraordinary precision. Photomasks are supplied either by captive mask shops within large fabs or by specialist merchant suppliers (CSET, 2025).

**EU Position:** The EU has no competitive position in the photomask market. The photomask market is dominated by Japan (~50%), the United States (~40%), and Taiwan (~7%), with some negligible share taken by South Korea and China. Notable suppliers include Dai Nippon Printing and Hoya (Japan); Photronics, Intel, and GlobalFoundries (US); TSMC and Taiwan Mask Corporation (Taiwan); Samsung (South Korea); and Shenzhen Newway and SMIC (China) (CSET, 2025).

#### #6.5 Photoresists

**Description:** Photoresists are chemicals deposited on wafers that selectively dissolve when exposed to patterned light passing through a photomask, forming the circuit pattern. Etching then transfers this pattern permanently onto the wafer. Photoresists are highly specific to the lithography process type such as EUV, ArF immersion, KrF, and others (CSET, 2025).

**EU Position:** The EU holds only a marginal position in photoresists. Japan dominates with firms like Shin-Etsu Chemical, JSR, Tokyo Ohka Kogyo (TOK), and Fujifilm Electronic Materials, controlling over 70% of the global photoresist market and even 95% for high-end EUV resists specifically (TrendForce, 2025e). Germany's Merck KGaA supplies photoresists and patterning materials globally, but the EU's overall market position remains small relative to Japan. The US (e.g. DuPont), South Korea (e.g. Dongjin Semichem), and China (e.g. Kempur, Nata Opto-electronic) hold the remaining market share (Fortune Business Insights, 2026).

#### #6.6 Resist Processing Tools

**Description:** Resist processing tools (commonly called “tracks”) coat wafers with photoresist using spin-coating, bake them to prepare the chemical layer, and develop them after exposure by chemically dissolving the patterned portions. Global market size: USD 3.4 billion (2024) (CSET, 2025).

**EU Position:** The EU has a small but real presence through Germany's SUSS MicroTec, owning approximately 2.9% of global market share. Japan dominates the market overwhelmingly with 93.8% overall market share, generated almost entirely by Tokyo Electron (TEL), and notably is the sole producer of tracks for advanced EUV lithography. Other players hold minor shares: South Korea's SEMES (~1.3%), US firm Lam Research (~1.1%), and China (less than 1%) (CSET, 2025).

#### #6.7 Etch and Clean Tools

**Description:** After photolithography creates a pattern in the photoresist, etch tools transfer that pattern into the permanent layer below. Cleaning tools then remove unwanted residual material. Etching and cleaning tools are each divided into dry and wet tools. Global market size: USD 24.7 billion (2024) (CSET, 2025).

**EU Position:** The EU has a very limited presence in etch and cleaning tools. Austria, Germany, and the Netherlands together occupy less than 1% of the global market share. Instead, the market is largely dominated by the US, through firms like Lam Research and Applied Materials (~50%), and Japan, through firms like Tokyo Electron, Screen, and Hitachi (~38%). China holds a growing share, including through NAURA and AMEC (~8%), while South Korea takes a smaller portion through SEMES (~3%), among other minor players (CSET, 2025).

#### #6.8 Process Control Tools

**Description:** Process control tools monitor wafers, photomasks, and the overall chip manufacturing process to ensure quality and consistency, making them among the most technically demanding and valuable equipment categories in semiconductor manufacturing. They encompass photomask inspection and repair tools, process monitoring tools, and wafer inspection and metrology tools. Global market size: USD 13.4 billion (2024) (CSET, 2025).

**EU Position:** The EU has a limited but notable presence in process control tools. The Netherlands accounts for ~5.2% of global market share through ASML's role in overlay metrology, and Germany contributes a further ~0.7% through Carl Zeiss. The market is otherwise overwhelmingly dominated by the US through KLA Corporation (~57%) and Applied Materials (~10%), and Japan through Lasertec (~8%) and Hitachi (~6%), with Israel, South Korea, and other countries accounting for the remainder (CSET, 2025).

#### #6.9 Wafer and Photomask Handlers

**Description:** Wafer and photomask handlers are robotic systems that store and transport wafers and photomasks within and between fabrication tools. These systems must operate in ultra-clean environments with extreme precision to prevent contamination or damage. Global market size: USD 2.4 billion (2024) (CSET, 2025).

**EU Position:** The EU is absent from this segment. The market is led by Japan owning ~86% of global market share through firms like Rorze, Muratec, and Daifuku, alongside South Korea's SEMES (~13%) and Taiwan suppliers (~1%) (CSET, 2025).

#### #6.10 Ion Implanters

**Description:** Ion implanters embed dopant atoms into silicon wafers to give different regions different levels of semiconductivity, enabling the creation of functional transistors. Tool variants serve different use-cases: high-current implanters for higher throughput; high-voltage implanters for deeper implants; and ultra-high-dose systems for very high dopant concentrations. Global market size: USD 3.5 billion (2024) (CSET, 2025).

**EU Position:** The EU is absent from ion implanter production. The United States dominates with ~90% global market share through the firms Applied Materials and Axcelis Technologies. Japan occupies ~9% of the global market through Sumitomo Heavy Industries Material Solutions and Nissin Ion, with Taiwan and China playing minor and emerging roles (CSET, 2025).

#### #6.11 Electronic Gases

**Description:** Electronic gases are used widely throughout semiconductor fabrication and packaging[3], serving varying purposes. These include common gases like nitrogen, hydrogen, helium, and argon, as well as specialty electronic gases like nitrogen trifluoride, tungsten hexafluoride, ammonia, and disilane, although EUV lithography equipment uses tin droplet laser-produced plasma rather than gas-based light sources (CSET, 2025).

**EU Position:** EU companies have a significant presence in electronic gases, led by France's Air Liquide and Germany's Linde, alongside other EU suppliers such as Solvay, Merck, Messer, and BASF. Other major non-EU suppliers include US firms Air Products and Entegris; Japanese firms like TNSC; and South Korea's SK Materials and Wonik, along with several Chinese players (CSET, 2025).

#### #6.12 Wafers

**Description:** Silicon wafers are ultra-pure silicon substrates that serve as the physical foundation on which chips are manufactured. While silicon dominates, some wafers also use alternative substrate materials (CSET, 2025).

**EU Position:** The EU has a minor share of the global wafer market, notably through Germany's Siltronic (~14%). Japan dominates the market, occupying ~56% market share through SUMCO and Shin-Etsu, followed by Taiwan's GlobalWafers (~16%), South Korea's SK Siltron (~10%), and many minor Chinese players (CSET, 2025).

#### #6.13 Wet Chemicals

**Description:** Wet chemicals like sulfuric acid, isopropyl alcohol, ammonium hydroxide, phosphoric acid, nitric acid, hydrochloric acid, and acetic acid are widely used in semiconductor manufacturing (CSET, 2025).

**EU Position:** The EU holds some market share, primarily through Germany's BASF and Belgium's Solvay. The rest of the market is dominated by Japan (e.g. Fujifilm, Kanto Chemical, Stella Chemifa), the US (e.g. Avantor, Honeywell), South Korea (e.g. Dongjin Semichem), and China (e.g. JHM, Runma, Sinyang) (CSET, 2025).

#### #6.14 CMP Materials

**Description:** Chemical mechanical planarisation (CMP) is a process that flattens the layers deposited during semiconductor fabrication, enabling lithography on each subsequent layer. The two primary consumables are chemical slurries, often incorporating rare-earth cerium oxide particles, and polishing pads, against which the wafer is pressed and rotated during planarisation (CSET, 2025).

**EU Position:** The EU holds a limited position in CMP materials, primarily through France's Saint-Gobain and Germany's Merck KGaA (CSET, 2025). The slurry market is led by the US and Japan: Fujifilm Electronic Materials holds ~31% (2024) of the slurry market, followed by Entegris in second place (Davis, 2024). In polishing pads, the US dominates with DuPont alone accounting for over 50% (2024) of the global CMP pad market. Other notable countries include Japan and China (Davis, 2024).

[3] Electronic gases are consumed across both Fabrication and Assembly, Testing and Packaging. However, to avoid duplication, they are listed here only once, under Fabrication (#6).

#### #6.15 CMP Tools

**Description:** CMP tools are the equipment used to perform chemical mechanical planarisation, a process in which layers deposited during semiconductor fabrication are flattened to enable accurate lithography at each subsequent step. A wafer is placed on a polishing pad with chemical slurry, and a rotating polishing head planarises the wafer surface (CSET, 2025).

**EU Position:** The EU has a negligible presence in the CMP tools market. The United States dominates through Applied Materials with ~60% market share, followed by Japan's Ebara Corporation with ~28%. China's Hwatsing (~9%) and South Korea's KCTech (~1%) occupy most of the remaining market (CSET, 2025).

### #7 Assembly, Testing, and Packaging

#### #7.1 Assembly Tools

**Description:** Assembly tools convert processed wafers into packaged semiconductor devices. This category includes four main sub-categories: (i) dicing tools, which separate individual dies from the wafer; (ii) bonding tools, including die attach, wire bonding, and advanced interconnect technologies, such as hybrid bonding, critical for leading-edge packaging; (iii) assembly inspection tools; and (iv) integrated assembly tools that combine multiple packaging steps. (CSET, 2025)

**EU Position:** The EU's presence in assembly tools varies considerably by sub-category. Its strongest position is in bonding tools, where the Netherlands' Besi holds ~19% global market share and Germany a further ~2%, though the segment is led by Singapore (~39%) and contested by South Korea (~19%), Japan (~14%), and China (~4%). The EU also holds a minor but notable share in assembly inspection tools through Germany's Mühlbauer (~2%), a market otherwise dominated by the US (~59%), Singapore (~22%), South Korea (~9%), and Taiwan (~6%). The EU is absent from dicing tools—nearly entirely Japanese-dominated (~95%)—and from integrated assembly tools, where Singapore (~98%) and the US (~2%) account for virtually the entire market. (CSET, 2025)

#### #7.2 Packaging Materials

**Description:** Semiconductor packaging requires several material inputs to bond a fabricated chip to a protective package: (i) bond wires (aluminium, copper, silver, or gold) that connect the chip to a lead frame or substrate; (ii) lead frames and substrates, which transfer signals between the chip and external devices; (iii) die attach materials (polymers and eutectic alloys) that attach the chip to packages or substrates; and (iv) ceramic packages, plastic substrates, or encapsulant resins that may be bonded to a chip. (CSET, 2025)

**EU Position:** The EU holds a notable position in select packaging material sub-categories through Germany. Heraeus is a leading global supplier of bond wires, competing directly with Japanese and Chinese firms. Germany's Henkel is also a significant supplier of die attach materials and encapsulation resins. The EU is, however, absent from both substrates, dominated by Japan, China, Taiwan, and South Korea, and lead frames, where China, Singapore, and Japan account for virtually the entire market. (CSET, 2025)

#### #7.3 Fabrication Tools for Advanced Packaging

**Description:** Advanced packaging tools are specialised semiconductor manufacturing systems used to integrate multiple chips into a single package, improving performance and power efficiency. This category includes tools for (i) CMP, (ii) deposition, (iii) etch/clean, (iv) inspection, (v) lithography, and (vi) wafer bonding and aligning. Global market size: USD 5.9 billion (2024). (CSET, 2025)

**EU Position:** The EU has a small but notable presence. Its strongest position is in wafer bonding, where Austria's EV Group (EVG) holds ~56% global market share, making it the world leader in this sub-segment, followed by Japan's Tokyo Electron (~39%) and Germany's SUSS MicroTec (~4%). SUSS MicroTec also holds a notable share in lithography tools for advanced packaging (~11%), a market otherwise led by Tokyo Electron (~66%) and US firm Applied Materials (~11%). However, the EU's presence in other sub-categories of fabrication tools for advanced packaging is either negligible (~1%) or completely absent, with the US, Japan, China, Israel, and Singapore collectively dominating (CSET, 2025).

#### #7.4 Packaging Tools

**Description:** Packaging tools encase and label dies in protective packages in conventional (non-advanced) packaging processes. This includes, for example, molding, trimming, forming, and marking equipment for encapsulating chips in plastic or ceramic packages (SKhynix, 2023). Global market size: USD 675.9 million (2024) (CSET, 2025).

**EU Position:** The EU has a small but notable presence. The Netherlands, through its firm Besi, occupies ~15% global market share, placing it as the third biggest supplier of packaging tools worldwide behind Japan (~58%) with firms like TOWA and APIC Yamada, and South Korea (~16%) with firms like HAMNI, EO Technics, and Genesem. Other countries like Singapore (~4%), China (~3%), and Taiwan (~1%) own rather minor shares. (CSET, 2025)

#### #7.5 Handlers and Probers

**Description:** Handlers and probers are equipment used during semiconductor testing. Probers make electrical contact with individual dies on a wafer for testing before packaging, while handlers transport packaged chips through test systems during final test. Global market size: USD 2.2 billion (2024). (CSET, 2025)

**EU Position:** The EU holds a minor presence, primarily through Italy's SPEA (~2%). The market is otherwise dominated by Japan (~44%), with firms such as Tokyo Electron, Accretech, and Advantest; South Korea (~22%), with firms such as Semics, SEMES, and TechWing; the US (~12%), with firms such as Cohu and FormFactor; China (~8%), led by Hangzhou Changchuan Technology; and smaller shares held by Taiwan (~5%), Singapore (~2%), and Malaysia (~2%). (CSET, 2025)

#### #7.6 Test Tools

**Description:** Test tools, also known as automated test equipment (ATE), verify the electrical functionality, performance, and reliability of semiconductors at both the wafer sort and final packaging stages. Sub-types include: (i) burn-in test tools, used for stress testing at elevated temperatures; (ii) linear and discrete test tools, for analog ICs and power semiconductors; (iii) memory test tools, for memory chips; and (iv) system-on-a-chip (SoC) test tools, covering a variety of processors including advanced logic chips. Global market size: USD 6 billion (2024). (CSET, 2025)

**EU Position:** The EU holds a negligible position in the ATE market, with Italy accounting for approximately 1% of global market share. The market is heavily concentrated between Japan's Advantest (~56%) and the US's Teradyne (~28%). Smaller shares are held by China's Hangzhou Changchuan Technology (~4%) and AccoTEST (~2%), South Korea's YIKC (~2%), and Taiwan's Chroma ATE (~2%). (CSET, 2025)

### #8 Compute Infrastructure

**Description:** Data centres are facilities that house the servers, networking equipment, and storage systems required to process, store, and deploy digital services, and increasingly to train and serve AI models at scale (McKinsey, 2025). Hyperscale data centres have emerged as essential AI infrastructure because they combine massive computing power, high-speed networking, and large-scale data storage in a form factor suited to web-scale workloads (OECD, 2025a). This layer is strategically important because compute is the physical backbone on which frontier AI development and deployment depend (European Commission, 2025e). Globally, demand is growing rapidly and data centre capacity is projected to triple by 2030, implying annual growth of roughly 19% to 27% (OECD, 2025a).

**EU Position:** The EU holds a weak position in compute infrastructure relative to the United States and China (European Commission, 2025e). Three US hyperscalers account for around 65% of the EU cloud market (Marcelin, 2025), whereas Europe's cloud sector is fragmented with only a handful of clouds like IONOS and OVHclouds exceeding EUR 1 billion in market capitalisation (Bria et al., 2025, pp. 69–70). Importantly, European industry has argued that hyperscalers' offers of sovereign government cloud amount to "sovereignty washing", masking continued extraterritorial control and data exposure behind rebranded products (Caffarra et al., 2025, p. 8). The picture is even starker for supercomputing[4], where the EU accounts for only 4.8% of global supercomputer performance, against 74.4% for the US and 14.1% for China (Pilz et al., 2025b). In response, France and Germany initiated GAIA-X in 2019 to develop a federated EU cloud infrastructure, though its impact remains contested (Scott & Bonfiglio, 2024). The European Commission subsequently approved IPCEI-CIS in 2023, authorising seven member states to mobilise up to EUR 1.2 billion in public funding for developing a federated and interoperable cloud to edge continuum (European Commission, 2023d). Most ambitiously, Europe's AI Continent Action Plan, released in April 2025, aims to triple EU data centre capacity within five to seven years and establish up to five AI Gigafactories, each with a computing capacity of 100,000 H100-equivalents (European Commission, 2025e).

[4] See footnote 1 above.

## #9 Data Centre Networking

**Description:** Data centre networking connects compute servers, storage systems, and management services within a data centre (Cisco, n.d.) and provides the high-bandwidth, low-latency connectivity needed to scale AI workloads across many nodes (NVIDIA, n.d.-b). Within AI data centres, networking is typically split between front-end networks, which handle storage, management, and external connectivity, and back-end networks, which connect accelerated servers for large-scale training and inference (HPE Greenlake, 2024). For AI back-end networks specifically, Dell'Oro Group forecasts that spending on switches deployed in accelerated server clusters will surpass USD 100 billion by 2030 (Dell'Oro Group, 2026).

**EU Position:** The EU holds a limited position in AI data centre networking. While it has emerging suppliers such as Nokia, it lacks a leading vendor comparable to US firms such as NVIDIA, Arista, and Cisco, or to Huawei in China. In frontier AI clusters, two fabric families dominate: InfiniBand, where NVIDIA remains the central supplier and which has long been the benchmark for large-scale AI training, and Ethernet, which offers greater openness and is rapidly gaining ground in hyperscale AI deployments (TrendForce, 2025c). By Q3 2025, Ethernet had already accounted for more than two-thirds of AI back-end switch sales, underscoring that the competitive centre of gravity is shifting toward Ethernet-based AI fabrics led primarily by US and Chinese firms rather than European ones (Dell'Oro Group, 2025).

## #10 Interconnects

**Description:** Interconnects are the high-speed, low-latency fabrics that bind multiple accelerators within a single server or rack into a unified compute unit, enabling them to share memory and exchange data at speeds that conventional interfaces such as PCIe cannot approach (ApX Machine Learning, n.d.). Unlike scale-out networking, which connects nodes across a cluster, scale-up interconnects operate across short physical distances, and prioritise memory coherence and near-zero latency above all else (NADDOD, 2025). Highly capable interconnects are critical for frontier AI, because communication overhead between GPUs is a major bottleneck in large-scale AI workloads (Alvarez et al., 2025).

**EU Position:** The EU has no presence in scale-up interconnects, and the market is highly concentrated around a single US company. The scale-up interconnect ecosystem for frontier AI systems is tightly integrated around NVIDIA's (US) proprietary offerings, spanning its NVLink interconnect protocol, NVSwitch fabric chips, and the NCCL communications library that optimises performance across all layers. No European company participates at the design level in any of these components, nor does any EU firm feature among the US-dominated founding members of the UALink Consortium, the main open-standard challenger to NVLink, although a minor fraction are participating as supporting members (e.g. ExpectedIT GMBH in Germany). The degree of lock-in is further compounded by the tight integration of NVLink with NVIDIA's broader software ecosystem, including NCCL and CUDA.

### #11 Network Infrastructure

**Description:** Networking infrastructure transmits data between data centres, users, and devices, forming the connective tissue of the digital economy and a prerequisite for AI. It includes access networks (e.g. 4G/5G, moving towards 6G); subsea and terrestrial fibre-optic systems that carry the bulk of long-distance traffic (OECD, 2025c); metropolitan optical networks that aggregate traffic within and between cities (Sousa de Sousa & Drummond, 2022); the internet core (IP routing, transit, and internet exchange points) that interconnects networks (Weller & Woodcock, 2013); and satellite links, including low-Earth-orbit (LEO) constellations, which extend connectivity to underserved and remote areas (OECD, 2025c).

**EU Position:** The EU retains important strategic assets in network infrastructure, particularly through Nokia and Ericsson, but its overall position is weakened by market fragmentation, lower investment intensity, and heavy competition from both the US and China. As of 2023, Huawei led the global telecom equipment market with around 30% market share, followed by Nokia and Ericsson at roughly 16% each, and ZTE at 10%. Europe's telecom operator market remains far more fragmented than those of its main competitors, with 34 mobile network operators compared with three in the United States and four in China, limiting economies of scale and constraining investment. This gap is reinforced by lower 5G population coverage with only 81% in the EU versus over 95% in the US and China, and by spectrum auction models that have often prioritised fiscal revenues over network deployment. In satellite connectivity, European operators remain focused on legacy MEO and GEO systems that cannot match the scale and responsiveness of US-based Starlink. In response, Europe initiated IRIS², a multi-orbital constellation of 290 satellites to provide sovereign secure connectivity. In subsea cables, limited and highly concentrated repair capacity creates an additional strategic vulnerability. (Council of the European Union, 2023, p. 4; Draghi, 2024, pp. 69–73; European Commission, 2024; Bria et al., 2025, pp. 61–62)

### #12 Consumer/Endpoint Devices

**Description:** Consumer and endpoint devices, such as smartphones, PCs, wearables, IoT systems, medtech devices, and industrial sensors, serve as the primary interfaces for user interaction with AI systems and are the primary sites for personal data aggregation. The strategic importance of these devices is increasing as the shift of inference workloads toward local compute reduces dependence on centralised cloud infrastructure (Stewart et al., 2025). The edge AI hardware market is projected to expand from USD 26 billion in 2025 to 59 billion by 2030 (MarketsandMarkets, n.d.). Smartphones are expected to account for 80.5% of edge AI hardware volume, with inference comprising 99.8% of workloads at the edge. Hundreds of millions of devices equipped with AI chips were sold globally in 2025 (Stewart et al., 2025).

**EU Position:** The European Union lacks a significant presence in consumer device manufacturing, operating systems, and app distribution, reflecting a broader pattern of concentrated global market control: Apple and Samsung alone accounted for approximately 39% of global smartphone shipments in 2025 (IDC, 2026). In 2024, Europe's smartphone market shipped 136 million units, with leading brands including Samsung, Apple, Xiaomi, Motorola, and HONOR, none of which are headquartered in the EU (Oyedeji, 2025). This creates a structural dependency, as European AI applications, data flows, and user interactions operate on hardware, operating systems, and app stores controlled by foreign entities. To address this, the EU deployed the Digital Markets Act (DMA) to reduce gatekeeper power and improve market contestability, requiring designated non-European platforms to open their ecosystems through interoperability, data portability, and prohibitions on self-preferencing, creating fairer conditions for European businesses. However, the EU retains meaningful strengths in industrial and medical endpoint devices. Bosch and Siemens are global leaders in industrial IoT (Bria et al., 2025, p. 63), with Germany's industrial IoT market projected to reach EUR 12.5 billion by 2029 (Statista, 2026). In connected medical devices, Philips (Netherlands) and Siemens Healthineers (Germany) rank among the top five global players, which collectively account for 65% of a USD 66.6 billion market (Global Market Insights, 2025).

## 4.3 Software

### #13 System Software and Frameworks

**Description:** System software and frameworks convert hardware capabilities into practical AI computation. Proprietary parallel computing platforms, particularly NVIDIA's CUDA, supply the drivers, compilers, libraries, and communication tooling essential for nearly all advanced model training. AMD's largely open source ROCm stack provides an alternative, though it has not yet reached full parity with CUDA across all workloads and deployment environments. Intel's oneAPI/DPC++, based on SYCL, a Khronos open standard for heterogeneous programming, represents a significant vendor-neutral programming model effort. Portability efforts include ONNX for model exchange and compiler ecosystems such as OpenXLA and MLIR, while projects like Triton provide more open approaches to GPU kernel programming. However, CUDA-native pipelines remain the prevailing standard.

**EU Position:** The EU currently lacks a significant domestic presence in system software and AI frameworks. NVIDIA dominates the data centre GPU market, supported by a developer community of over 5.9 million CUDA developers built over two decades (Fernandez, 2025; NVIDIA, 2025). The ecosystem lock-in associated with CUDA, which encompasses thousands of optimised applications, remains unmatched by European alternatives, leaving a critical AI component dependent on US-controlled ecosystems (Fernandez, 2025). Open compiler solutions such as OpenXLA and MLIR improve portability, but do not yet match CUDA's ecosystem depth, backend coverage, and developer lock-in for frontier-scale workloads. The European Chips Act, together with the European Processor Initiative (EPI) and the DARE programme built upon it, targets the full software-hardware stack, extending European strategic ambition beyond silicon to interoperable system software. The AI Continent Action Plan with the EuroHPC AI Factories initiative establishes a foundation for open-stack deployment: AI Factories are designed to provide a rich software environment including ready-to-use AI-oriented tools, clear use-cases for efficient large-scale use, and support for containerised workloads and workflows (EuroHPC JU, 2024).

## #14 Algorithms and Models

**Description:** Core innovation in AI occurs at the algorithmic and model layers. Transformer architectures and diffusion models form the foundation of nearly all advanced general-purpose and multimodal AI systems. Industry has further consolidated its dominance in model development. According to the Stanford AI Index (2025), nearly 90% of notable AI models in 2024 originated in industry, up from 60% the previous year, even as academia remains the primary source of highly cited research. Open-weight models have closed the benchmark performance gap with closed systems from 8% to 1.7% within a single year (Maslej et al., 2025, p. 95).

**EU Position:** The EU ranks second globally in AI-related scientific publications, but it produced only three notable AI models in 2024, all originating from France, compared to forty developed in the United States (Maslej et al., 2025, p. 46). Mistral AI, the leading frontier model developer in the EU, achieved a valuation of EUR 11.7 billion in September 2025 and remains the sole EU-based developer consistently monitored among global frontier model producers (Djurickovic, 2025). Beyond Mistral, the EU is also pursuing a consortium-based approach through OpenEuroLLM, a 20-partner project to develop open source, multilingual foundation models (OpenEuroLLM, 2025). Despite this strong research base, EU firms struggle to convert scientific output into scalable frontier technologies, as the innovation pipeline weakens at the commercialisation stage due to fragmented public and private R&D funding, shallow later-stage capital markets, and weaker productivity returns on R&D investment compared to the United States (Manganelli, 2025, pp. 3–4). The European Frontier AI Initiative, launched by France, Germany, and the European Commission in November 2025 as the EU's best-funded non-profit frontier AI programme, seeks to address these shortcomings by consolidating computational resources, talent, and long-term funding to develop sovereign, open frontier models at scale.

## #15 Applications and Services

**Description:** The applications and services layer translates model capabilities into operational products such as API-based developer tools, AI-native software-as-a-service (SaaS) platforms, vertical-specific software for sectors such as healthcare, finance, and legal, AI coding assistants, enterprise copilots, and agentic workflow tools. Some of these AI-enabled applications can have significant sovereignty effects, such as smart ad bidding systems. This layer operates above the model hosting and inference infrastructure, transforming raw model capabilities into user-facing functionality via product interfaces, prompt orchestration, fine-tuning integrations, and domain-specific data pipelines. Spending on AI-enabled applications is projected to grow faster than any other segment through 2029, with service providers accounting for 80% of infrastructure spend as they scale to support massive increases in agentic workloads, and enterprises accelerate the shift from pilots to production deployments (IDC, 2025).

**EU Position:** Application access is primarily mediated by operating systems, app stores (Apple App Store, Google Play Store), and cloud orchestration layers (Amazon Web Services (AWS), Google Cloud, Microsoft Azure, Cloudflare), which are predominantly controlled by a limited number of non-EU platform operators. North America accounts for over 40% of global AI SaaS revenues, while EU providers represent only a minor share (Market Data Forecast, 2026). European players such as SAP, Mistral AI, and vertical AI startups in healthcare and manufacturing are active in this layer, but remain subscale relative to US incumbents. Major application markets, including enterprise productivity, online advertising, and application distribution, are predominantly controlled by US-based platforms (Sukharevsky, 2024). This dependency is compounded by lock-in effects: once AI workflows are embedded, switching costs become prohibitive, and enterprises prefer extending existing vendor relationships over adopting new entrants (Challapally et al., 2025, pp. 15, 18). Europe's reliance on non-EU platforms is identified as a structural competitive weakness, with the EU failing to translate its innovation strengths into commercially viable digital technologies, and its disadvantage in cloud and platform services is likely to widen (Draghi, 2024, pp. 67–68). EU policy responses focus on strengthening competition and market access, particularly by implementing interoperability and data portability requirements outlined in the Digital Markets Act (DMA) and the Data Act. Additionally, the AI Act introduces regulatory sandboxes that enable companies to test innovative AI systems under supervisory oversight.

## 4.4 Data

### #16 Generic/Open

**Description:** Web-crawled text, open source code, imagery, satellite feeds, and government databases form the foundational corpus layer on which frontier AI models are trained. The major open web repository Common Crawl alone contains approximately 130 trillion tokens, the indexed web around 510 trillion, and the full web an estimated 3,100 trillion (Maslej et al., 2025, p. 59). Yet, if current AI scaling trends continue, then the effective stock of public human-generated data might be exhausted between 2026 and 2032 (80% confidence) (Villalobos et al., 2024). Failures at this component, through legal restrictions, linguistic gaps, or corpus depletion, propagate directly into model capability ceilings.

**EU Position:** EU AI developers face a relatively restrictive and administratively complex environment for web-scale training data. Under the DSM Copyright Directive, commercial text-and-data mining is permitted only where rights holders have not expressly reserved their rights, while the AI Act requires GPAI providers placing models on the EU market to implement additional policies for compliance with Union copyright law and related transparency obligations (Marcelin & Cassetti, 2025). The GDPR further imposes substantial compliance obligations on AI models whenever personal data is involved (European Data Protection Board, 2024). The Data Union Strategy and digital omnibus are intended to ease these bottlenecks by increasing data availability and simplifying the EU data-rule landscape, although some critics caution that simplification may weaken the EU's existing safeguards on privacy, enforcement, and fundamental rights (EDRi, 2025).

### #17 Specialised/Proprietary

**Description:** Sector-specific datasets—like electronic health records, genomics, financial data, legal corpora, industrial process data, and defence-related data—provide the domain-specific signal that generic web corpora cannot, and access to such high-volume, high-quality, and specialised datasets has become a defining factor of global competitiveness in AI (European Commission, 2025m, p. 3). Unlike open corpora, specialised datasets are often not publicly accessible and are shared only under sector-specific legal and governance conditions, which can restrict access and reuse and, in some cases, make them strategically sensitive for AI development (European Commission, n.d.-m; European Commission, 2025m, pp. 1–4).

**EU Position:** The EU holds valuable sectoral and specialised data assets, but many remain siloed or underused, limiting their reuse for AI development (European Commission, 2025m, p. 1). Between 2021 and 2024, the European Commission invested EUR 336 million in 14 strategic common European data spaces, which now need to be scaled across the EU (European Commission, 2025m, p. 2). The EU also plans to establish data labs linking data holders, data spaces, and AI ecosystems to support data pooling, curation, and reuse for AI (European Commission, 2025m, p. 4–5). Separately, industry-led platforms such as Manufacturing-X and Catena-X illustrate how trusted sector-specific data sharing can be operationalised in practice.

### #18 Synthetic

**Description:** Synthetic data is artificially generated to reproduce the statistical properties, patterns, and relationships of real datasets (European Commission, 2025m, p. 5). Its principal advantages include privacy preservation, mitigation of data scarcity, cost reduction, and enhanced scenario modelling, positioning it as an increasingly significant input to AI development (World Economic Forum, 2025, pp. 3–7). In specialised settings such as self-play and formal reasoning, it has already proved highly effective, as demonstrated by AlphaZero and AlphaProof (Sevilla et al., 2024). However, synthetic data currently functions as a complement rather than a general substitute for real-world data, and its potential to resolve data-scaling bottlenecks at scale remains unproven (Sevilla et al., 2024). It also carries material risks, including bias amplification, degraded model accuracy, and in recursive training settings, model collapse (World Economic Forum, 2025, p. 8).

**EU Position:** The EU is beginning to build coordinated synthetic-data infrastructure, but remains considerably behind leading non-EU frontier labs. The Data Union Strategy commits the European Commission to develop guidance and interoperability standards for trusted synthetic-data use, proposes a dedicated synthetic data factory via EuroHPC, and envisages data labs that will support synthetic-data generation and related services for AI development (European Commission, 2025m, pp. 8, 11). However, US frontier labs are far ahead, already embedding synthetic data generation as a core component of their model development pipelines (Google DeepMind, 2025, p. 2; OpenAI, 2025b), whereas current EU efforts are still focusing on establishing the enabling infrastructure.

### #19 Human Data Labour

**Description:** Human input remains integral to the AI data pipeline across dataset collection, curation, annotation, content moderation, and reinforcement learning from human feedback. This work is predominantly outsourced to lower-income economies (notably Kenya, the Philippines, and India), where per-task rates are lowest and labour protections least developed. The resulting geographic concentration of data labour raises systemic concerns about worker welfare and exploitation that current regulatory frameworks have yet to adequately address (Open Data Institute, 2025, pp. 6–12).

**EU Position:** The EU is beginning to build an EU-governed alternative to parts of the offshore annotation model through the data labs proposed in the Data Union Strategy, which will provide curation, labelling, annotation, enrichment, (pseudo)anonymisation, and vectorisation services for AI development (European Commission, 2025m, pp. 6–8). The strategy also announces a dedicated initiative to standardise annotation and labelling practices (European Commission, 2025m, p. 12), while the Platform Work Directive gives the EU a stronger framework on algorithmic management and platform-worker protections than many competing jurisdictions. By contrast, leading Chinese and US labs often outsource large annotation workloads to lower-income economies primarily for cost reasons, sustaining a structural cost and scale advantage (Dosunmu & Waithira, 2025) that EU-based alternatives may struggle to match under stronger labour and governance standards.

### #20 Deployment Data Flywheel

**Description:** A deployment data flywheel is a self-reinforcing loop in which data collected from deployed AI interactions is continuously used to refine AI models, generating better outcomes and more valuable data for further improvement. As systems are deployed at scale, they capture inference logs, user feedback, and real-world usage patterns, which feed back into cycles of data processing, model customisation, and redeployment, progressively improving accuracy, adaptability, and cost efficiency over time (NVIDIA, n.d.-c).

**EU Position:** The EU is structurally weaker in this layer because it lacks consumer AI platforms at the scale to generate behavioural data compared to leading Chinese and US incumbents. In early 2026, OpenAI (US) recorded over 700 million weekly active users (OpenAI, 2026), while DeepSeek (China) reached around 100 million in April 2025 (Backlinko, 2026). By contrast, the strongest public usage metric for the EU's leading native contender, Mistral's Le Chat, is only 1 million downloads in the 14 days after its February 2025 launch (Caruso, 2025), with no comparable public active-user disclosure.

## 4.5 Enabling Environment

### #21 Talent

**Description:** AI development requires diverse human expertise throughout the development and deployment lifecycle. This includes semiconductor engineers and data centre specialists at the hardware layer, ML engineers and data scientists for model training and optimisation, and researchers, algorithm designers, and governance professionals working at the frontier. These roles exhibit significant differences in scarcity. Three proficiency tiers structure the talent landscape: AI-literate workers (Tier 0), software and data professionals (Tier 1), and advanced AI researchers and engineers (Tier 2), with shortfalls concentrated at Tier 0 and Tier 2 (Pal et al., 2025, p. 8).

**EU Position:** The European Union's talent position remains structurally precarious despite a robust academic foundation. Europe maintains a 30% higher per-capita concentration of AI professionals than the United States and nearly three times that of China, with 70% of these professionals holding a Master's degree or PhD (Pal, 2024). In contrast, the US uses a strong retention model, with over 80% of internationally trained AI PhDs staying in the country after graduation (Zwetsloot et al., 2019). China is rapidly expanding domestic AI talent pipelines by prioritising state-led recruitment and investing heavily in AI education (Gehlhaus et al., 2022). Despite this per-capita advantage, Europe faces persistent shortages of ICT specialists (European Commission, 2025g, p. 4), and a net outflow of senior AI professionals to the United States, primarily due to the presence of leading US universities, a massive tech sector, and more efficient immigration processes (Pal, 2024). Among major European countries, France experiences a net loss of AI professionals, while Germany, although it attracts regional talent, also loses substantial numbers to the United States and the United Kingdom (Pal, 2024, p. 9). At the advanced level, Tier 2 talent constitutes only 15% of AI talent on average across European countries, against 24% of AI vacancies, pointing to a structural undersupply of advanced AI expertise (Pal et al., 2025, p. 28). Structural brain drain to the United States persists, as AI salaries in the US are 30% to 70% higher than in comparable European positions (Kiorri, 2026). In response, the European Union's AI Continent Action Plan addresses both talent supply and retention through initiatives such as the AI Skills Academy, AI fellowships, returnship programmes, and streamlined visa pathways under the revised Blue Card Directive.

### #22 Financial Resources

**Description:** Frontier AI development is exceptionally capital-intensive. The amortised cost of training the most compute-intensive models has increased by 2.4 times annually since 2016 (Cottier et al., 2025, pp. 1, 5). Epoch AI estimates that the largest training runs will exceed USD 1 billion by 2027 (Cottier et al., 2025), with US hyperscalers already having committed USD 443 billion in capital expenditure in 2025 (Chalfin & Pugh, 2025). Sustaining this level of investment requires three concurrent financial streams: private finance (including venture capital, corporate research and development, and hyperscaler capital expenditure), public finance (such as research grants, development bank instruments, and public procurement), and blended finance (public-private partnerships and subsidised compute access).

**EU Position:** The EU faces a structural, widening capital gap. In 2025, US firms attracted 75% (USD 194 billion) of global AI venture capital, while the EU27 secured only 6% (USD 15.8 billion) (OECD, 2026b, p. 8). Mega-deals exceeding USD 100 million accounted for 73% of total AI investment value, a funding scale the EU rarely achieves (OECD, 2026a). In contrast, the European Commission reports that the EU invests only 4% of the amount the United States allocates to AI overall (European Commission, 2025b). The EU's structural disadvantage in venture capital extends beyond what headline figures indicate. The EU accounts for only 5% of global VC funds raised, compared to 52% in the US and 40% in China. VC investment represents just 0.05% of EU GDP, nearly six times lower than in the US and UK, with the largest gap in later-stage financing (Draghi, 2024, p. 242). In response, the European Commission launched InvestAI (with a EUR 200 billion target, including EUR 20 billion for AI Gigafactories) and is advancing the Savings and Investments Union.

### #23 Policy and Governance

**Description:** Policy and governance instruments, including AI-specific legislation, semiconductor frameworks, cybersecurity mandates, and international coordination, establish the legal environment for the development, access, and auditing of frontier AI systems. The EU AI Act is the Union's core AI-specific framework, while the Digital Services Act (DSA) and the Digital Markets Act (DMA) govern adjacent risks linked to platforms, distribution, and market power in the digital environment in which AI systems are developed and deployed. This framework applies to all providers whose AI systems and models are accessible to EU users, irrespective of the provider's place of establishment. At the international level, coordination is facilitated through the OECD and Global Partnership on Artificial Intelligence (GPAI) member countries integrated partnership (Perset & Aranda, 2025).

**EU Position:** The European Union is the first global actor to implement comprehensive AI regulation, anchored by the EU AI Act which covers both AI systems and general-purpose AI (GPAI) models. The broader regulatory stack, which includes the GDPR, the Data Act, and the Data Governance Act, further shapes the AI landscape by governing data flows, access conditions, and privacy standards essential to AI systems. The Digital Omnibus, enacted in November 2025, postpones the enforcement of standalone high-risk obligations until December 2027. As of August 2025, 14 member states had not designated a national competent authority, resulting in a lack of functional enforcement mechanisms and imposing regulatory costs estimated at EUR 150 billion annually on businesses, falling disproportionately on SMEs (Dzurinda et al., 2025, p. 17). The EU AI Office has exclusive enforcement authority for frontier AI but is constrained by structural capacity limitations. By comparison, the United States lacks binding federal AI legislation, relying instead on sectoral regulation and executive guidance, while China embeds AI governance within state-led industrial policy, prioritising strategic deployment over independent oversight (Chan et al., 2025). On the international stage, the European Union promotes its regulatory model through the 'Brussels to the Bay' programme. The GPAI Code of Practice, endorsed by Amazon, Anthropic, Google, Microsoft, and OpenAI, is emerging as a potential global standard for oversight of GPAI models. To address regulatory fragmentation in the single market for AI, the European Commission has proposed the 28th regime. This optional EU-wide legal framework would enable companies to operate under one coherent set of rules, regardless of their member state of incorporation.

### #24 Standards

**Description:** Standards translate regulatory principles into concrete and testable requirements, including those related to AI safety, reliability, and interoperability. The European Commission has explicitly framed standard-setting as soft power: standards shape which technologies can be deployed and who sets the global rules, carrying "an indirect bearing on the EU's economic security" (European Commission, 2023b, p. 9). These standards enable AI providers to demonstrate compliance and allow regulators to verify adherence. For the AI Act to be operational, the European Commission must publish harmonised standards in the Official Journal. Once published, these standards provide providers with a presumption of conformity with the Act's requirements for high-risk systems. Two bodies develop AI standards relevant to the European Union: (i) CEN-CENELEC JTC 21, the EU's standardisation body comprising over 300 experts from more than 20 countries and (ii) the international ISO/IEC JTC1/SC42 committee. The alignment of the latter with the US NIST AI RMF influences whether EU standards facilitate or restrict global market access.

**EU Position:** The European Union possesses strong institutional standard setting capacity, yet it has not met its self-imposed deadlines. CEN-CENELEC has indicated that the complete standards package will not be available before the fourth quarter of 2026. The Digital Omnibus, published in November 2025, aims to address this issue by making the enforcement of high-risk AI systems, as defined under the EU AI Act, contingent upon the availability of harmonised standards, establishing a final deadline of December 2027 (Niestadt, 2026). The delay is partly attributable to structural factors. A Joint Research Centre analysis found that ISO/IEC 42001:2023 does not align with the product-safety framework of the AI Act, necessitating the development of a tailored standard, prEN 18286, which entered public enquiry only in October 2025. In the absence of published harmonised standards, conformity assessment processes cannot advance, and the EU's capacity to establish credible global benchmarks remains limited. Globally, AI standards are shaped primarily through ISO/IEC JTC1/SC42, in which EU member states participate as voting members, while the secretariat and chair positions are held by the United States. The EU's structural advantage is its ability to link harmonised standards directly to binding law through the AI Act, a mechanism not present in the United States, which relies on the voluntary NIST AI RMF, or in China, which coordinates through state-directed bodies.

### #25 Societal Context

**Description:** The societal context of AI includes public attitudes, AI literacy (defined as the ability of citizens and workers to critically engage with AI systems), and patterns of enterprise AI adoption. As AI deployment accelerates, the gap between societal readiness and technological capability presents a structural sovereignty risk. This gap influences the extent of democratic oversight and determines whether Europe can harness AI-driven change. Cultural norms related to privacy, safety, and accountability, together with public trust in companies, emerging technologies, and expectations concerning fundamental rights, shape both the pace of AI deployment and the level of political support.

**EU Position:** Public support for AI oversight in the European Union is substantial, with 78% of citizens considering it important for authorities to shape AI development (European Commission, n.d.-e). Currently, only 55.6% of adults in the European Union possess basic digital skills (European Commission, n.d.-j), compared to the Digital Decade target of 80% by 2030 (European Commission, 2023c). Enterprise AI adoption reflects a similar gap: only 19.95% of EU companies currently use AI, with especially midcaps and SMEs struggling with AI adoption, a large gap against the EU's own 2030 Digital Decade target of 75% AI take-up in enterprises (Eurostat, 2026; European Commission, 2023c). According to Cedefop's 2024 AI skills survey, six out of ten employees are susceptible to AI-driven task transformation (Cedefop, 2024). The Digital Education Action Plan (2021–2027) and a December 2025 Eurobarometer shows 92% of citizens support the inclusion of digital skills in school curricula, demonstrating growing momentum (European Commission, 2025p). The EU has responded with binding obligations, including the Article 4 AI literacy obligation of the AI Act, which mandates that providers and deployers ensure sufficient AI literacy among their staff and users. Expanding AI literacy campaigns that target both citizens and workers will be essential to bridging the gap between societal readiness and the rapid pace of AI deployment. By contrast, China has integrated AI literacy into national school curricula as a strategic priority, with approximately 61% of Chinese adults reporting at least basic digital skills, while the United States is pursuing an executive mandate for AI education in K–12 schools, albeit through a decentralised and unevenly resourced system (The State Council of the People's Republic of China, 2024; Toczauer, 2025).

### #26 Market Dynamics

**Description:** Market dynamics in frontier AI are shaped by the interaction of supply-side concentration, investment flows, market concentration, and geopolitical and industrial intervention. Supply-side concentration is dense across many AI supply chain components like chip design, cloud provision, and frontier model development, arising from capital intensity, economies of scale, proprietary software ecosystems, vertical integrations, and switching costs that entrench incumbents (OECD, 2025b, pp. 21–28). High concentration further shapes market contestability: the OECD observes that "once a firm secures a dominant position, the incentives to innovate may shift, potentially leading [to] firms seeking to protect their position" through exclusionary conduct, conditional rebates, or killer acquisitions (OECD, 2025b, p. 24). Finally, geopolitical and industrial policy dynamics constitute a structurally distinct driver: AI's significance to national competitiveness and security has led to increasing governmental intervention across the supply chain, with trade barriers and export controls distorting markets, fragmenting supply chains, and entrenching dominant players (OECD, 2025b, pp. 29, 31).

**EU position:** The EU occupies a structurally weak position across many dimensions of frontier AI market dynamics. On supply-side and market concentration, Nvidia holds over 80% of the global AI accelerator market, TSMC manufactures 99% of the world's AI accelerators, and AWS, Microsoft Azure, and Google Cloud command over 60% of the global cloud market (OECD, 2025b, pp. 13–14, 17), with the EU holding meaningful positions only in select upstream segments such as EUV lithography equipment. Investment flows compound this disadvantage with US firms having attracted 75% (USD 194 billion) of global AI venture capital in 2025, while the EU27 secured only 6% (USD 15.8 billion) (OECD, 2026b, p. 1, 8). On geopolitical dynamics, US export controls constrain EU supply chain options without reciprocal industrial benefit. While the EU's Chips Act, InvestAI initiative, and AI Continent Action Plan provide meaningful industrial instruments, their impact remains limited: the Chips Act is projected to deliver only 11.7% of global semiconductor manufacturing share by 2030 against a 20% target (European Court of Auditors, 2025, p. 7), and the AI Gigafactories envisaged under the AI Continent Action Plan represent only a fraction of buildout plans by foreign competitors (Epoch AI, n.d.-b).

# 5. A Unified Framework

**This section introduces and applies a unified framework for assessing EU frontier AI sovereignty.** We proceed in four steps. First, we connect the five sovereignty pillars developed in Section 3 to the Frontier AI Stack mapped in Section 4, establishing a single analytical lens through which policy interventions can be assessed against their sovereignty implications (Section 5.1). Second, we provide a comprehensive map of 92 EU frontier AI policy initiatives across four major European Commission communications (Section 5.2). Third, we apply the framework as a backward-looking diagnostic to five selected initiatives, illustrating how it surfaces opportunities, risks, and cross-pillar trade-offs that current policy debates tend to leave implicit (Section 5.3). Finally, we demonstrate a more thorough application of the framework through a detailed assessment of the EU's AI Gigafactory initiative, showing how the six standardised questions can structure a comprehensive sovereignty analysis of a single flagship intervention (Section 5.4).

## 5.1 Introducing the Framework

**The framework (Figure 4) connects Europe's sovereignty pillars to the Frontier AI Stack, illustrating how frontier AI sovereignty—understood as control over, or influence across, specific components of this stack—serves as a critical enabler of the Union's broader goals.** Rather than treating frontier AI sovereignty as an end in itself, the framework positions it as a means: the enabling condition through which Europe can pursue a more holistic conception of sovereignty, extending well beyond narrow economic gains. Concretely, the framework links frontier AI to five pillars of European sovereignty: (i) **Economic Competitiveness**; (ii) **Resilience and Preparedness**; (iii) **Security and Defence**; (iv) **European Values and Identity**; and (v) **Foreign Relations and Perceptions.**

**Each layer, component, and sub-component of the Frontier AI Stack impacts every sovereignty pillar, though to varying degrees.** Critically, an intervention in the stack can advance one pillar while undermining another, making cross-pillar trade-off analysis essential to any policy assessment. By decomposing the Stack into five layers comprising 26 components and 29 sub-components, the framework equips policymakers to identify precisely where interventions are needed, and where leverage points lie.

**The framework makes policy objectives more nuanced and trade-offs explicit.** For example, policies aimed at Economic Competitiveness may focus on expanding domestic data centre capacity and access to high-quality data to support frontier AI research and deployment. Yet large-scale compute imposes high energy demands with direct implications for Resilience and Preparedness, as growing data centre load can erode grid reserve capacity and reduce the margin available to absorb shocks during emergencies. Similarly, the use of extensive datasets to train economically competitive AI models raises trade-offs with European commitments to data protection, privacy, and individual rights. Frontier AI sovereignty, in this sense, is not about maximising any single objective, but about retaining the capacity to navigate such trade-offs deliberately.

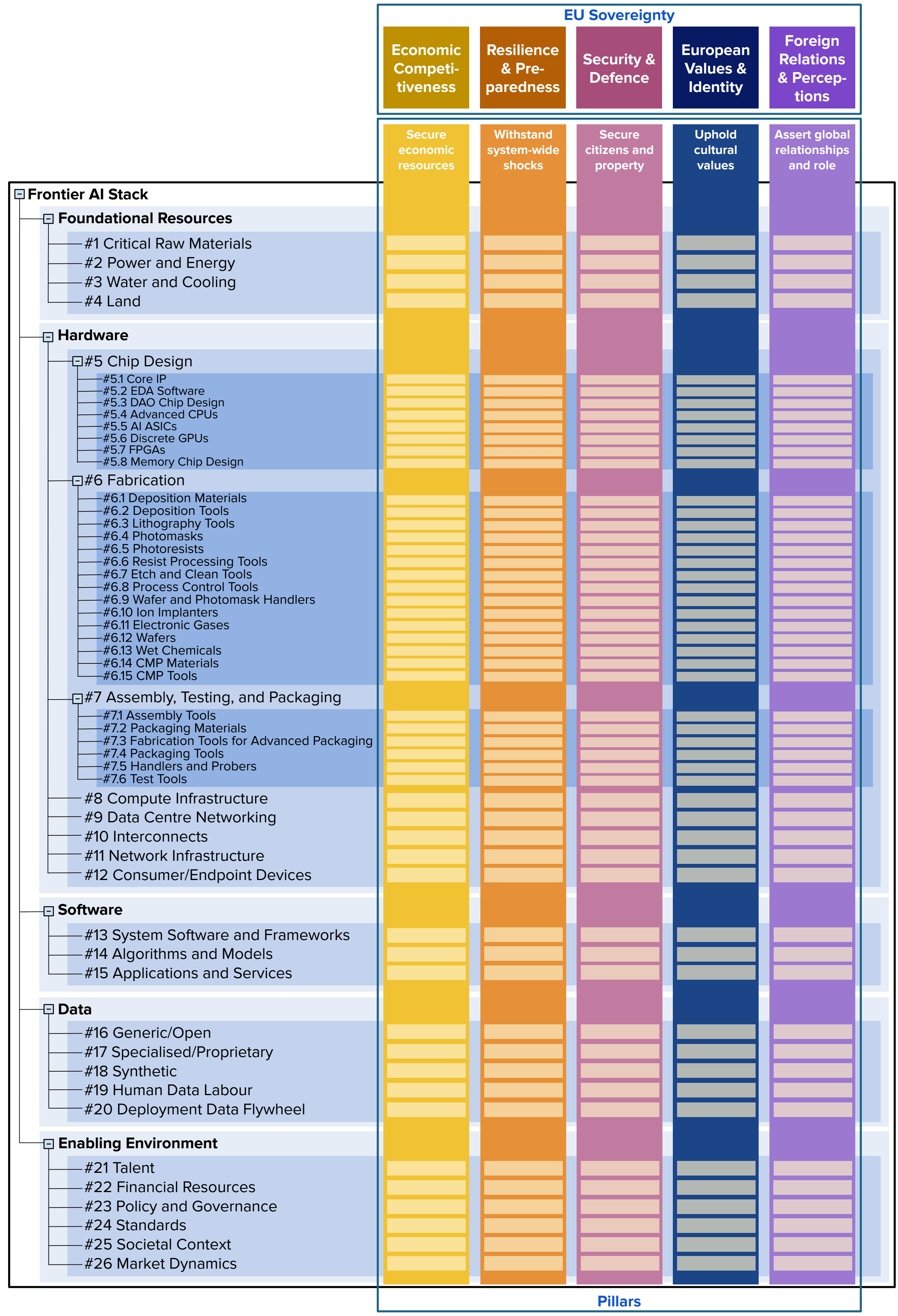

***Figure 4:*** *A unified framework for frontier AI sovereignty, connecting a granular view of the frontier AI stack to five fundamental pillars of European sovereignty. Sovereignty is treated as a multi-dimensional condition. Interventions in the stack can have positive and negative effects across pillars, making trade-offs explicit and policy objectives more nuanced.*

## 5.2 Mapping Existing EU AI Policy Initiatives

**The following section maps the EU's existing AI policy portfolio (see Appendix A.1 for the full list), providing the empirical foundation against which the framework's diagnostic value is demonstrated in Sections 5.3 and 5.4.**

**A systematic comparative analysis of four major EU communications uncovers 92 distinct policy initiatives which characterise the EU's AI policy portfolio.** The EU AI initiatives dataset was constructed using a structured data-annotation process applied to four primary European Commission communications (or frameworks): (i) COM165 (AI Continent Action Plan), (ii) COM723 (Apply AI Strategy), (iii) COM724 (Resource for AI Science in Europe [RAISE]), and (iv) COM835 (Data Union Strategy). Data extraction was performed by a single annotator, and validation was subsequently conducted by an additional team member to ensure accuracy and consistency. Data collection was deliberately limited to information explicitly provided in the relevant European Commission communications. Where budgetary figures and timeline details were not specified in the primary documents, targeted supplementary searches were conducted to identify publicly available budget allocations. This approach yielded a comprehensive dataset of 92 initiatives, providing a structured foundation for analysing the EU AI initiative landscape.

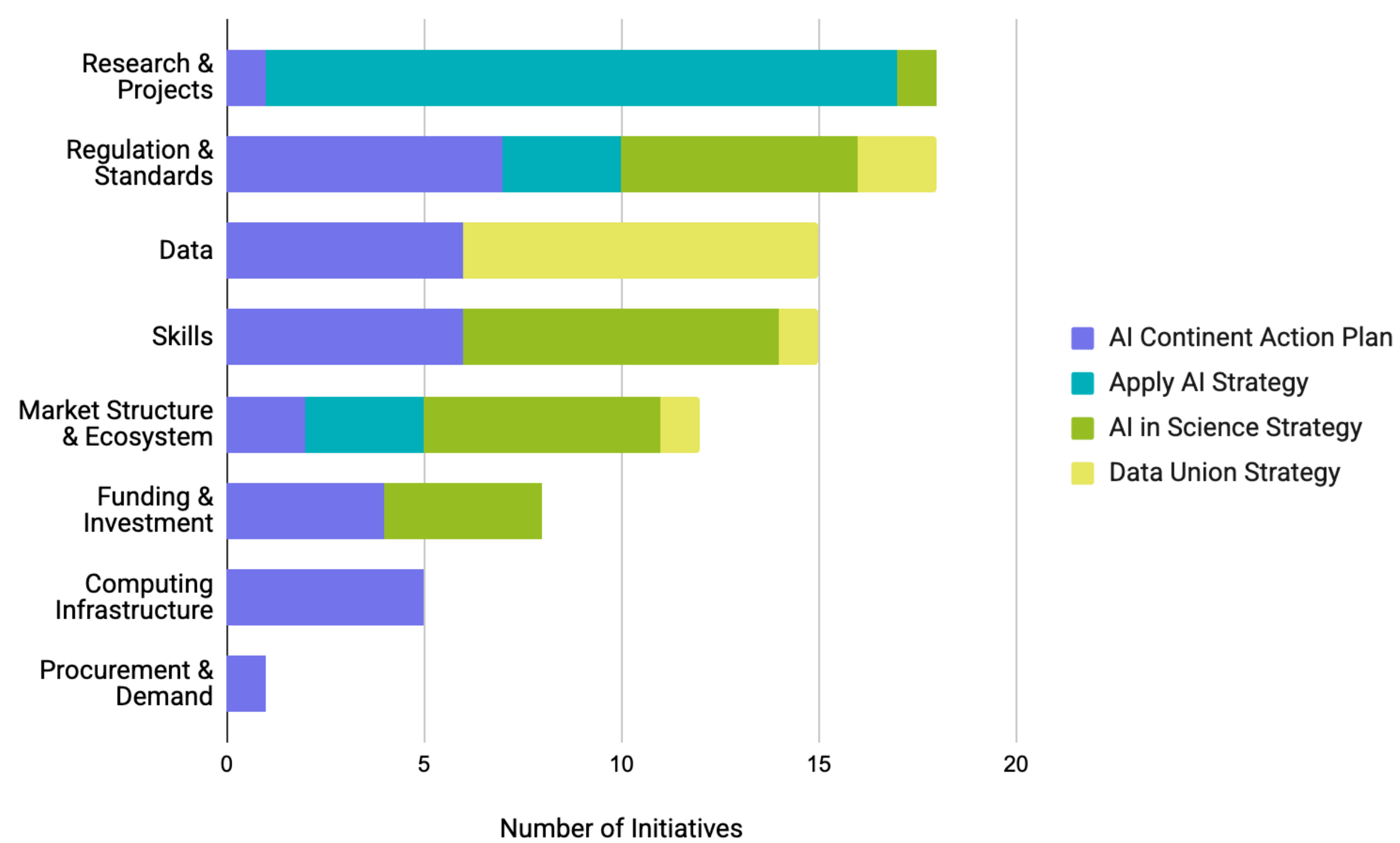

***Figure 5:*** *Distribution of initiatives by thematic categories and European Commission communication.*

**To provide a first descriptive overview of the portfolio, initiatives are grouped into eight thematic categories.** Five of these reflect the pillars identified in the AI Continent Action Plan: Data, Skills, Research and Projects (Development of Algorithms and Adoption in COM165),

Regulation and Standards (Regulatory Simplification in COM165), and Compute. The remaining three—Market Structure and Ecosystem, Funding and Investment, and Procurement and Demand—capture development functions that recur across the documentation but are not formally named as pillars in COM165. Together, these eight categories form the taxonomy used in Figure 5.

**These categories do not neatly map onto the layers of the Frontier AI Stack.** This gap is itself informative: it indicates that the attention of the European Union is more concentrated on certain layers of the stack (such as compute or data), while other layers receive comparatively little structured attention. This uneven coverage supports our hypothesis that a holistic analysis of the frontier AI stack is lacking.

**The distribution across the eight thematic domains highlights a strong emphasis on governance, innovation, and foundational enablers.** The twin emphasis on Regulation and Standards and Research and Projects, each comprising 18 initiatives and representing the largest categories, demonstrates the European Commission's commitment to both establishing clear regulatory guidelines and promoting innovation. This balance exemplifies the European Union's approach to advancing the deployment of AI within a comprehensive regulatory framework. The prominence of Data and Skills, each represented by 15 initiatives, demonstrates that Europe's competitiveness relies on fostering other frontier AI stack components such as high-quality datasets and human capital, rather than on deployment alone. The Market Structure and Ecosystem category, with 12 initiatives, further underscores coordinated efforts among national actors and the strengthening of innovation networks across Europe. While Computing Infrastructure comprises only five initiatives, these include flagship projects such as AI Gigafactories and the Cloud and AI Development Act, indicating that strategic priority is not solely determined by the quantity of initiatives.

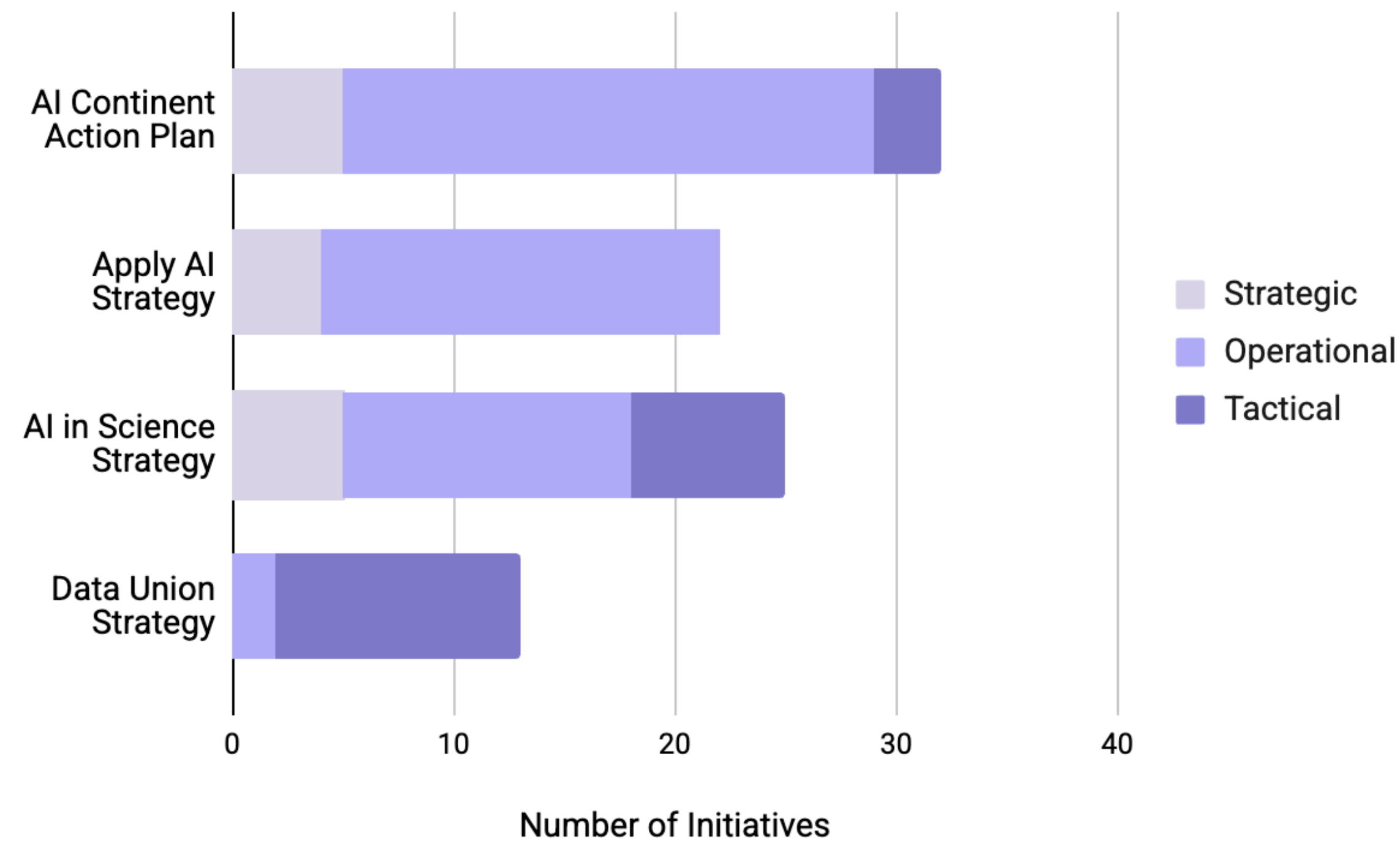

***Figure 6:*** *Distribution of initiatives by strategic horizon.*

**We clustered the 92 initiatives across three strategic horizons.** Interventions were classified as Strategic, Operational, or Tactical, depending on whether a combination of the implementation, scope, and effects were expected in a more or less extended period of time.

**The initiative portfolio is primarily focused on implementation: the Operational cluster accounts for 57 of 92 initiatives (62%).** Such a focus highlights the EU's priority on delivery and near-term capability development, such as EuroHPC Supercomputers, which actively builds the shared computing infrastructure underpinning EU AI capacity. The Tactical cluster comprises 21 initiatives (23%) that offer short-term, responsive support for early deployment and enforcement; for instance, the AI Act Service Desk provides immediate regulatory guidance to businesses navigating compliance. The Strategic cluster includes 14 initiatives (15%) that define medium- to long-term plans, especially in areas needing sustained investment such as skills, data, and computing infrastructure, such as the Horizon Europe Investment Agenda, which commits sustained, multi-year research and innovation funding to build Europe's long-term AI capabilities.

**Budget information is largely absent: 62 initiatives (67%) have undetermined budgets.** Only 16 initiatives (17%) have clearly specified budgets, 11 (12%) are components of broader programmes without separate budget lines, and 7 (7%) operate under regulatory mandates that do not require dedicated budgets. Where budgets are specified, allocations vary considerably. For example, AI Factories are allocated EUR 10 billion for 2021 to 2027; InvestAI targets EUR 200 billion in mobilised investment, combining EUR 50 billion in public funding with EUR 150 billion from private investors, including a dedicated EUR 20 billion fund for AI gigafactories; GenAI4EU receives EUR 700 million; and EIC AI Support is allocated between EUR 150 million and EUR 400 million.

**Geographic coverage is primarily pan-European, with 61 initiatives (66%) operating across the EU.** Among the remaining initiatives, three distinct patterns emerge: (i) 10 initiatives (11%) have global or international components, (ii) 4 initiatives are limited to specific member states, usually those with AI Factories or pilot sites, and (iii) 2 initiatives provide remote access for member states lacking direct infrastructure. Geographically targeted initiatives include AI Factory Antennas, serving member states without factories; the Alliance for Language Technologies, covering 17 member states; and Autonomous Drive Ambition Cities, focusing on selected EU cities. Global-facing initiatives comprise programmes that attract international researchers through the MSCA "Choose Europe" Scheme and the Private Sector Pledging Campaign, which targets both EU and global participants.

**Within this policy landscape, several flagship initiatives stand out as foundational building blocks for Europe's AI sovereignty agenda, operating primarily across four layers and components in the stack:** compute infrastructure (#8), algorithms and models (#14), data (#16, #17, #18, #19, #20), and policy and governance (#23).

1. **On compute infrastructure,** the AI Gigafactories and Cloud and AI Development Act anchor Europe's push for computational sovereignty, reinforced by AI Factories, EuroHPC Supercomputers, and the AI Factory Antennas network, which together form a distributed infrastructure backbone from advanced training capacity to regional access points.

2. **On data,** the Data Union Strategy establishes 14 sector-specific Common European Data Spaces, with Data Labs bridging these spaces to AI ecosystems, and Copernicus and the European Open Science Cloud providing open access to scientific and Earth observation data.
3. **On algorithms and models,** the Frontier AI Initiative—a public-private partnership launched by France, Germany, and the European Commission in November 2025, with an Expert Forum convening on 23 April 2026—is supported by RAISE and the Network of Frontier AI Labs on the research side, and GenAI4EU on the demand side.
4. **On policy and governance,** the AI Act Service Desk, AI Regulatory Sandboxes, the GPAI Code of Practice, and the Apply AI Alliance collectively support compliant and coherent AI deployment across the single market, with particular attention to easing compliance for SMEs and startups.

## 5.3 Applying the Framework to EU Frontier AI Policy

**The EU's 92 AI policy initiatives catalogued in Section 5.2 reflect genuine political momentum, but it is rarely explicit which sovereignty objective a given initiative is designed to advance, or how it relates to others in the portfolio.** Investments in compute infrastructure, regulatory frameworks for foundation models, and data-sharing schemes all touch on sovereignty, but without a common analytical language it is difficult to assess whether they address the right pressure points, whether their effects reinforce or undermine one another, and whether the portfolio as a whole holistically advances European sovereignty.

**This is precisely what the sovereignty framework from Section 5.1 is designed to address.** By mapping specific interventions onto the constituent layers of the Frontier AI Stack and assessing their impact across the five sovereignty pillars, it makes explicit the objectives and policy trade-offs involved. The framework is designed for dual application: as a tool for forward-looking strategic planning, it helps derive targeted interventions in the AI stack from a vision for advancing one or several sovereignty pillars; as a diagnostic instrument, it enables a backward-looking assessment of existing initiatives.

1. **A forward-looking application finds and prioritises interventions in the frontier AI stack and proceeds through five steps:** (i) assessing the EU's current position across frontier AI stack components; (ii) deriving, from the EU's sovereignty objectives, what each pillar requires from specific stack components; (iii) identifying measures to close the gap between the EU's current position on the stack components and its strategic requirements; (iv) analysing the opportunities, risks, and cross-pillar trade-offs of those measures; and (v) prioritising interventions against one another and against alternative uses of the same resources.
2. **A backward-looking application starts from existing policy rather than strategic objectives, and proceeds through four steps:** (i) mapping the initiatives relevant to the analysis; (ii) identifying which frontier AI stack components each initiative primarily affects; (iii) examining Europe's current position on the stack components, (iv) assessing, for each initiative and its targeted stack components, which opportunities, risks, and cross-pillar trade-offs it could generate across the five sovereignty pillars, using relevant pillar-specific KPIs; and (v) drawing on these assessments to identify how existing initiatives could be strengthened or recalibrated to better advance the EU's sovereignty objectives, and what gaps might remain for improving frontier AI sovereignty. This is the approach we adopt in the remainder of this section, particularly in 5.4 when we assess the EU's AI Gigafactories proposal.

**Below, we illustrate the framework's backward-looking application by assessing five initiatives drawn from the 92 identified in Section 5.2.** A full forward-looking application—deriving new interventions from the EU's sovereignty vision across the entire stack—requires strategic planning and threat assessment exercises that are beyond the scope of a public report. A backward-looking assessment, however, can be demonstrated directly against the EU's existing policy portfolio that we mapped in Section 5.2, making it better suited to illustrate the framework's analytical value. The five initiatives each target different primary Frontier AI Stack components, spanning compute infrastructure (#8), financial resources (#22), algorithms and models (#14), specialised/proprietary data (#17), talent (#21), and policy and governance (#23), illustrating the framework's applicability across different stack layers. The analysis is intentionally illustrative: for each initiative, we examine only the single primary stack component identified, and the opportunities, risks, and trade-offs outlined are exemplary rather than comprehensive, without a systematic application of pillar-specific KPIs. A deeper application to the AI Gigafactories initiative follows in Section 5.4.

| Policy Intervention | Primary Stack Component | Analysis Lens | Sovereignty Pillars | | | | |
|---|---|---|---|---|---|---|---|
| | | | Economic Competitiveness | Resilience and Preparedness | Security and Defence | European Values and Identity | Foreign Relations and Perceptions |
| **AI Regulatory Sandboxes:** Controlled environments under Article 57 of the AI Act in which providers can develop and test innovative AI systems under regulatory supervision before they are placed on the market or put into service. | **#23 Policy and Governance** | **Opportunities** | Accelerate time to market for novel solutions, where regulatory requirements are yet to be defined, and support compliance for SMEs with limited resources. | Help institutions to anticipate and prevent risks through evidence-based regulatory learning. | Pre-market testing can surface security vulnerabilities before deployment. | Enable compliance testing of AI systems against EU-value anchored regulation. | Promote regulatory learning and the EU's AI governance and deployment expertise relative to the rest of the world while de-risking market entry for foreign start-ups. |
| | | **Risks** | Offer only surgical support to SMEs and in cases of regulatory uncertainty; risks crowding out pro-innovation policies at scale. | Test systems under controlled conditions that may not capture real-world failure modes under stress, creating a false sense of preparedness. | Foreign providers may engage in "sandbox shopping" and choose less stringent national sandboxes to siphon EU data with less oversight (Marcelin, 2026). | AI Act Art. 59's data reprocessing provisions allow personal data collected for other purposes to be used for AI development in a sandbox, stretching the GDPR. | Conservative sandbox design (continued liability, strict GDPR constraints) may alienate third-country partners, who may favour more permissive sandbox approaches. |
| | | **Selected Trade-offs** | 1. Accelerating sandbox approvals for **Economic Competitiveness** risks premature deployment, undermining **European Values and Identity** (unresolved fundamental rights risks) and **Resilience and Preparedness** (unresolved safety risks).<br>2. Imposing security requirements on participants (e.g. mandatory penetration testing) fosters **Security and Defence** but might reduce sandbox accessibility and speed, weakening **Economic Competitiveness**.<br>3. Extending sandbox knowledge-sharing to other jurisdictions might improve **Foreign Relations and Perceptions** but cause tensions with **Security and Defence** through potential technology leakage. | | | | |
| **InvestAI:** InvestAI mobilises EUR 200 billion in investment (EUR 50 billion public, EUR 150 billion private) to close | **#22 Financial Resources** | **Opportunities** | EUR 200 billion could meaningfully narrow the EU's structural capital gap with the US and strengthen its competitive position across key frontier AI stack components. | Reduces strategic foreign dependencies along critical frontier AI stack components through targeted, coordinated investment. | Boosts investments into EU security-relevant applications (e.g. intelligence analysis), currently often dominated or controlled by foreign providers. | The EUR 50 billion public component may be conditioned on advancing EU values (e.g. prioritising AI for health or education). | A well-funded AI sector could strengthen Europe's geopolitical standing in international trade and AI governance, beyond being a regulatory superpower. |
| the EU's structural AI capital gap across the full AI value chain, including a dedicated EUR 20 billion fund for AI Gigafactories. | | **Risks** | Even if fully mobilised, it remains modest in scale relative to US commitments (e.g. Stargate alone allocates USD 500 billion for compute infrastructure). | Public pressure to spend and limited availability of domestic alternatives may force investment in deals that exacerbate existing dependencies. | Long-term financial viability of security-domain initiatives depends on sustained government contracts that are not yet guaranteed, risking stranded investments if defence procurement cycles lag or spending priorities shift. | As citizens grow increasingly critical of AI, replicating US-style large-scale investments risks alienating the public and undermining societal cohesion. | A significant gap between the EUR 200 billion headline figure and actual mobilised capital would risk undermining EU credibility if the initiative fails to deliver on its commitments. |
| | | **Selected Trade-offs** | 1. Maximising private capital mobilisation for **Economic Competitiveness** requires offering attractive returns, yet attaching strong public-interest conditionalities to advance **European Values and Identity** (e.g. prioritising AI in health and education) may reduce private investor appetite.<br>2. Building redundant, EU-localised frontier AI supply chains for **Resilience and Preparedness** is capital-intensive and often cost-inefficient relative to global alternatives, potentially undermining **Economic Competitiveness**.<br>3. Prioritising rapid buildout of frontier AI stack components critical for **Resilience and Preparedness** may require relaxing **European Values and Identity**-aligned conditionalities like sustainability and safety requirements. | | | | |
| **Frontier AI Initiative:** Launched by France, Germany, and the European Commission in November 2025, the Frontier AI Initiative is a non-profit, public-private partnership to develop sovereign frontier AI models at scale, addressing the EU's widening frontier model | ***#14* Algorithms and models** | **Opportunities** | Pursuing neglected R&D approaches (e.g. world models) rather than replicating scaling could enable paradigm leapfrogging and position the EU as a first-mover in emerging AI product markets. | Reduces critical dependency on a small number of foreign model providers whose access may be restricted under geopolitical pressure. | Supports sensitive national security applications (e.g. military planning, intelligence analysis) while mitigating security and data leakage risks from foreign models. | Could prioritise safety-by-design model architectures which are uniquely positioned to protect European values (e.g. fundamental rights). | Serves as a springboard for international AI R&D partnerships, strengthening the EU's position as a collaborative science and technology actor. |
| | | **Risks** | Risks producing technically sophisticated but commercially irrelevant models, reducing the initiative to a prestige project with limited economic impact. | Concentrating resources in a small number of EU research efforts may weaken broader ecosystem resilience relative to a more distributed funding approach. | Pressure to deploy models in critical domains may outpace the development of adequate security protocols in comparably smaller EU frontier AI efforts. | Competitive pressure to match US/Chinese capabilities may incentivise capability racing at the expense of safety. | Framing the initiative in competitive terms against US/China risks provoking retaliatory measures (e.g. chip export restrictions) rather than enabling productive collaboration. |

<table>
<tr>
<td>gap with the United States and China.</td>
<td></td>
<td>Selected Trade-offs</td>
<td colspan="5">1. Economic Competitiveness may incentivise aggressive data collection, reduced safety testing, and compressed timelines, potentially directly conflicting with the governance standards central to Resilience and Preparedness, European Values and Identity, and Security and Defence.<br>2. Extending R&D partnerships to non-EU countries (e.g. UK, Canada) can strengthen Foreign Relations and Perceptions and Economic Competitiveness through access to talent and resources, but opens technology-sharing channels that elevate IP leakage risks, undermining Security and Defence.<br>3. A distributed, member state-led approach reinforces European Values and Identity but risks diluting resources below the threshold required for frontier-competitive performance, weakening Economic Competitiveness.</td>
</tr>
<tr>
<td rowspan="3">European Health Data Space: Creates an EU-wide framework for sharing and reusing electronic health data for healthcare, research, innovation, and policymaking. Aims to unlock fragmented health data through common interoperability rules and cross-border infrastructure. Key provisions apply from 2029.</td>
<td rowspan="3">#17 Specialised/ Proprietary</td>
<td>Opportunities</td>
<td>Could generate EUR 11 billion in savings over the next decade through enhanced data accessibility, and drive 20–30% expansion in the European digital health sector (European Commission, n.d.-n).</td>
<td>Cross-border health data interoperability could directly strengthen pandemic detection, monitoring, and response capacity (Regulation (EU) 2025/327 [EHDS Regulation], 2025).</td>
<td>Secondary-use access to population-scale clinical and genomic data could accelerate R&D on medical countermeasures against CBRN threats.</td>
<td>Could establish strong patient data rights, including seamless electronic access to health records, portability, rectification, and reversible opt-out from secondary use.</td>
<td>EHDS access could serve as a meaningful asset in international health research collaboration and trade negotiations.</td>
</tr>
<tr>
<td>Risks</td>
<td>Projected economic gains are contingent on solving significant techno-legal challenges, (e.g. interoperability) (van Drumpt et al., 2025).</td>
<td>A malfunction within the centralised EHDS infrastructure could trigger systemic failures across healthcare applications or pandemic response efforts.</td>
<td>Pooling and interconnecting health data at scale elevates exposure to cyber threats and systemic breach risk (van Kessel et al., 2023)</td>
<td>GDPR exemptions within the EHDS framework risk undermining rather than strengthening patient data rights (European Data Protection Board 2022).</td>
<td>Deploying EHDS access as a diplomatic lever risks being perceived as coercive data diplomacy, undermining the EU's standing as a rules-based rather than power-based actor.</td>
</tr>
<tr>
<td>Selected Trade-offs</td>
<td colspan="5">1. Strengthening privacy protections advances European Values and Identity but may reduce the volume and granularity of available health data, potentially constraining Economic Competitiveness and Resilience and Preparedness.<br>2. Mandating high cybersecurity standards for EHDS infrastructure reinforces Security and Defence but raises operational costs and may slow data access turnaround, reducing Economic Competitiveness.<br>3. While organising European health data in a single platform accelerates pandemic response and bolsters Resilience and Preparedness, it may create systemic cyber and bioengineering risks that undermine Security and Defence.</td>
</tr>
</table>

| **EU AI Gigafactories (AIGFs):** Aims to establish up to five large-scale compute facilities, each with a computing capacity of 100,000 H100-equivalents. Backed by a EUR 20 billion blended-finance facility under InvestAI, the Gigafactories are designed to provide sovereign compute infrastructure for frontier AI research, training, and deployment. | **#8 Compute Infrastructure** | **Opportunities** | Could enhance the competitive positioning of EU industries (e.g. frontier AI, pharma, robotics) and sovereign frontier AI model R&D through large-scale compute access. | Sovereign compute infrastructure reduces dependence on foreign cloud providers, whose access may be restricted under geopolitical pressure. | Sovereign and secure compute is a prerequisite for running sensitive security workloads without exposing data to foreign-controlled systems. | Confidential computing environments with attestation and logging could provide cryptographic assurances, enabling practical enforcement of the AI Act and other digital rules. | Sovereign compute ownership could constitute a bargaining asset for international negotiations. |
|---|---|---|---|---|---|---|---|
| | | **Risks** | If AIGFs fail to offer relevant compute at competitive prices, they may crowd out alternative AI investments with better returns. | AIGFs remain critically dependent on foreign components like NVIDIA chip design and TSMC fabrication, offering operational but not supply-chain resilience. | Concentration of sovereign compute in five known locations creates high-value physical and cyber targets for adversarial states. | Data centres' large ecological footprints risk conflicting with the EU's sustainability commitments. | Even at full buildout, planned capacity may be insufficient to confer meaningful geopolitical leverage given the scale of competing foreign infrastructure investments. |
| | | **Selected Trade-offs** | 1. Higher security postures foster **Security and Defence** but raise capital and operating costs, potentially elevating user prices and constraining **Economic Competitiveness.**<br>2. Full **Resilience and Preparedness** demands running all AI workloads on European soil, which, given higher energy and operational costs, may be less economically viable than selective reliance on foreign compute, undermining **Economic Competitiveness**.<br>3. Expanding sovereign data centre capacity may go against sustainability commitments, conflicting with **European Values and Identity**, even as it advances other pillars like **Resilience and Preparedness** and **Economic Competitiveness**. | | | | |

***Table 4:*** *Illustrative backward-looking application of the unified frontier AI sovereignty framework to five selected EU initiatives, each targeting a different primary stack component. For each initiative, the table outlines example opportunities and risks across the five sovereignty pillars and surfaces selected cross-pillar trade-offs.*

## 5.4 AI Gigafactories: A Detailed Assessment

**Having illustrated the framework's breadth across five diverse initiatives, we now apply it in greater depth to the EU's AI Gigafactory (AIGF) initiative.** This deeper assessment moves beyond the condensed pillar-by-pillar overview provided in Table 4 to demonstrate how the unified sovereignty framework can structure a more thorough, backward-looking analysis of a single flagship intervention. The assessment follows six standardised questions, applicable to any backward-looking use of the framework:

1. What is the intervention's objective?
2. Which Frontier AI Stack component(s) does this intervention primarily concern?
3. What is the EU's current position on this/these component(s)?
4. How does this intervention influence each sovereignty pillar?
5. What trade-offs arise across these pillars when pursuing the intervention?
6. How to improve the intervention?

As with the illustrative assessments above, a comprehensive application would require analysing all Frontier AI Stack components affected by the initiative: an exercise best left to the civil servants administering it who have internal and confidential knowledge of the programme. For tractability, we focus on the primary component concerned, compute infrastructure (#8), and we only include a few non-comprehensive and illustrative KPIs for each pillar from Section 3, adapted to this component.

**1. What is the intervention's objective?**

The EU wants to increase its compute infrastructure, as it considers this to be a crucial and limiting component to developing and deploying sovereign frontier AI. Advancements in frontier AI have been closely linked to the scaling of compute across pre-training, post-training, and inference. In other regions, like the US, this motivates rapid AI-dedicated data centre buildouts (e.g. OpenAI's Stargate). In response, the EU decided to expand its large-scale data centre capacity through the creation of up to five AI Gigafactories, each with a compute capacity of 100,000 H100-equivalents (European Commission, 2025e).

**2. Which Frontier AI Stack component(s) does this intervention primarily concern?**

AIGFs primarily concern the component compute infrastructure (#8). However, they are composed of and dependent on upstream components like chips (#5–7), data centre networking (#9), interconnects (#10), or power and energy (#2), and affect several downstream components, including the EU's ability to attract and retain talent (#21). As mentioned above, for illustrative purposes the analysis below solely focuses on implications on compute infrastructure (#8).

**3. What is the EU's current position on this/these component(s)?**

The EU's largest public computing facilities are currently its existing AI Factories, each hosting up to 25,000 H100-equivalents. While several private European compute providers also operate at scale (e.g. OVHcloud, Scaleway), none can match the capacity of the leading foreign hyperscalers (Amazon Web Services, Google Cloud, Microsoft Azure). Taken together, the EU accounts for only about 4.8% of global supercomputer performance in March 2025, far behind the United States at 74% and China at 14.1% (Pilz et al., 2025b).

**4. How does this intervention influence each sovereignty pillar?**

**Economic Competitiveness:**

**AIGFs could sustain sovereign AI R&D, deployment, and industrial competitiveness, while retaining cloud revenues that currently flow to foreign hyperscalers. Yet the planned scale is dwarfed by foreign buildouts, and potentially insufficient demand risks stranded assets.**

Opportunities: AIGFs could guarantee the compute access required to sustain sovereign AI R&D and deployment. By integrating high-performance compute with sectoral data pooling and specialised security architectures, these facilities can bolster specific industries such as healthcare and robotics. Granting SMEs and domestic industry preferential access can also create the strategic certainty for long-term investments in AI innovation and adoption. Lastly, AIGFs may create second-order effects, whereby the value currently absorbed by US cloud hyperscalers remains with European compute providers.

Risks: However, European compute demand may not reach the scale AIGFs are designed for, risking stranded assets (Warso, 2026). The EUR 20 billion allocation also represents a significant opportunity cost relative to investments into AI adoption or other demand-side areas that may yield more immediate competitiveness returns. Moreover, AIGF scale remains heavily outpaced by foreign buildout projects (e.g. OpenAI's Stargate alone has mobilised USD 500 billion), raising the question of whether 100,000 H100-equivalents per facility will still be sufficient for frontier AI competitiveness by the time these facilities become operational.

Potential KPIs: (i) EU share (%) of global AI supercomputing capacity; (ii) PPP-adjusted price differential for key compute inputs (e.g. GPU-hour) between the EU and comparable non-EU OECD countries; and (iii) share (%) of EU enterprises reporting domestic frontier-scale compute access as a factor in AI adoption decisions.

**Resilience and Preparedness:**

**AIGFs provide Europe with a sovereign compute fallback that cannot be withdrawn by foreign actors. Yet they remain vulnerable to upstream supply-chain dependencies and energy constraints on already strained European grids.**

Opportunities: Sovereign compute infrastructure provides the EU with a domestic fallback that cannot be withdrawn, sanctioned, or subject to pricing pressure by foreign providers during geopolitical crises, providing a concrete safeguard for critical public services and economic infrastructure currently dependent on US-controlled cloud. This capability grows more strategically significant as AI becomes more deeply embedded in key economic and public services. Furthermore, AIGFs distributed across multiple EU member states would create geographic resilience: if one facility is disrupted by an energy crisis or natural disaster, others can absorb critical workloads.

Risks: AIGFs remain critically dependent on foreign-controlled components across the stack, including NVIDIA GPUs, TSMC fabrication, and imported critical raw materials from countries such as China, Malaysia, and Russia. While sovereign compute built on non-sovereign components offers operational resilience (e.g. data stays in the EU), it does not guarantee full supply-chain resilience: a disruption at any upstream node (e.g. expanded US export controls) could leave AIGFs unable to procure replacement hardware. AIGFs also impose high, concentrated loads on local electricity grids which already are the binding constraint for data centre expansion in major EU hubs (Ember, n.d.), leaving both the facilities and the workloads they host vulnerable to disruption during energy crises.

Potential KPIs: (i) adoption rate (%) of EU public-sector and strategic-industry workloads migrated to sovereign compute within three to five years; and (ii) estimated recovery time (in months) required to restore pre-disruption service levels for critical functions (e.g. payments, cloud) following a foreign-cloud provider withdrawal scenario.

**Security and Defence:**

**Sovereign, secure compute is a prerequisite for credible EU defence AI, but concentrating capacity in a small number of facilities built on foreign hardware introduces new physical, cyber, and supply-chain vulnerabilities.**

Opportunities: Reserving a share of AIGF capacity as secure compute certified at SL3+ (as defined by Nevo et al. (2024)) could provide credible domestic infrastructure for running sensitive security workloads without exposing data to foreign-controlled systems. Such capacity could also enable the EU to develop frontier AI models end-to-end on secure European infrastructure without foreign processing at any stage. Beyond hosting, a dedicated allocation of AIGF capacity toward security-relevant military AI—in domains such as intelligence analysis and cyber defence—could materially strengthen the EU's defence capabilities.

Risks: However, concentrating sovereign compute in five publicly known, large-scale facilities creates high-value physical and cyber targets (Grunewald & Gershovich, 2025), where a successful attack on even one site could cause severe disruption to EU sovereign compute capacity (Amoah et al., 2025). Further, operations where civilian and military workloads share the same infrastructure heighten the risk of sensitive information leakage if security tiers are insufficiently segregated. Moreover, despite European ownership requirements, AIGFs will be built using overwhelmingly non-European hardware, raising the unresolved question of whether facilities constructed from foreign components can be considered genuinely secure against supply-chain attacks or hardware-level backdoors that operational security alone cannot mitigate (Grunewald & Gershovich, 2025).

Potential KPIs: (i) share (%) of EU security and intelligence AI workloads run, stored, and processed on EU-sovereign compute infrastructure; (ii) share (%) of those workloads running on EU-owned compute certified at SL3+ (RAND); and (iii) rate of successful breaches of AIGF infrastructure relative to comparable non-EU compute facilities.

**European Values and Identity:**

**AIGFs could strengthen EU regulatory enforcement and enable dedicated capacity for public-interest purposes. Yet their ecological footprint risks conflicting with sustainability commitments, and private capital may steer capacity rather towards commercial uses.**

Opportunities: Confidential computing environments with attestation and logging mechanisms could enable practical digital regulation enforcement, moving compliance from self-certification to cryptographic evidence that safety tests were executed within certified environments. Sovereign compute infrastructure also provides the EU with domestic alternatives to foreign services, potentially strengthening the EU's capacity to credibly enforce its values-based regulation on foreign technology providers. Public governance structures could further ensure that a share of AIGF capacity is deliberately allocated to public-interest purposes aligned with European values such as AI in education or health, rather than defaulting to purely market-driven allocation.

Risks: However, private investors contributing the majority of AIGF funding may steer operations toward commercially profitable workloads at the expense of public-interest applications. AIGFs' high ecological footprint also risks conflicting with EU sustainability commitments. If facilities are not powered by demonstrably renewable energy and operated under binding resource-efficiency standards, they might undermine the EU's commitment to addressing climate change risks.

Potential KPIs: (i) share (%) of frontier AI models which offer cryptographic evidence that safety tests were executed within certified environments; and (ii) share (%) of European compute allocated to public interest AI applications.

**Foreign Relations and Perceptions:**

**AIGFs could serve as bargaining assets in shaping international negotiations, collaborations, and AI governance. Yet this might depend on whether AIGFs achieve sufficient scale and utilisation compared to foreign buildouts.**

Opportunities: Sovereign compute ownership could constitute a tangible bargaining asset in international negotiations, as control over frontier-scale infrastructure increasingly shapes who sets terms in AI governance (Shamir & Negele, 2025). Further, by offering secure, high-capacity compute to like-minded democracies, AIGFs could anchor multinational AI development partnerships that pool resources across AI middle powers as a counterweight to the US–China duopoly (Martinet et al., 2025). The operational experience from building and maintaining exascale infrastructure would also deepen EU technical expertise, strengthening its positions in international AI standards bodies (e.g. ISO/IEC JTC 21, ITU-T) and governance forums (e.g. Hiroshima Process, UN AI Advisory Body), where technical credibility is a prerequisite for normative influence.

Risks: However, these geopolitical gains depend on AIGFs achieving sufficient scale and operational success. Even at full buildout, planned capacity may be insufficient to confer meaningful leverage given competing foreign investments like OpenAI's Stargate. Shamir and Negele (2025) instead propose a target of 15% of worldwide AI compute by 2030, which would require investment far exceeding EUR 20 billion. Further, if AIGFs are perceived internationally as underutilised or failed projects, they could actively damage the EU's credibility as a serious technological actor and weaken its standing in international governance forums. Moreover, framing AIGFs explicitly as a sovereignty play against US and Chinese compute dominance risks undermining the transatlantic partnerships the EU simultaneously depends on for critical AI stack components.

Potential KPIs: (i) number of active bilateral or multilateral AI initiatives to pool compute resources; and (ii) EU representation (%) in leadership and working-group positions across major AI governance bodies (e.g. GPAI, Hiroshima Process, UN AI Advisory Body).

**5. What trade-offs arise across these pillars when pursuing the intervention?**

Security and Defence vs. Economic Competitiveness: Higher security postures (e.g. SL3+) (Nevo et al., 2024) may result in greater capital and operating costs, potentially pricing out commercially promising use cases (Kumar et al., 2022). However, this premium should be weighed against unmet demand from sensitive sectors that currently avoid cloud adoption entirely. A tiered security architecture that is stringent for classified workloads, commercial-grade for others (Kothuri et al., 2025), could manage this tension at the cost of additional architectural complexity.

Resilience and Preparedness vs. Economic Competitiveness: Full Resilience and Preparedness would demand sovereign compute infrastructure to host all EU AI R&D and inference workloads. Yet higher EU energy costs and maintenance prices likely make this economically unattractive (Meyers, 2025) for workloads where advantages such as security and political stability cannot offset additional costs. Furthermore, building quickly requires foreign components, perpetuating the dependencies AIGFs are designed to reduce, while waiting for European alternatives (if they ever materialise) risks delays hampering competitiveness.

European Values and Identity vs. Economic Competitiveness: Large-scale compute infrastructure potentially strengthens competitiveness but imposes substantial energy demands (IEA, 2025)—costs that conflict with the EU's sustainability commitments. Binding environmental standards and renewable energy requirements can mitigate but not eliminate this tension. Siting decisions that prioritise sustainability may exclude economically optimal locations, while relaxing environmental criteria risks undermining the EU's climate leadership credibility.

**6. How to improve the intervention?**

AIGFs represent an important step toward closing Europe's gap in large-scale compute infrastructure. Yet an analysis of AIGFs across the sovereignty pillars reveals several areas for improvements to help AIGFs deliver on their promise.

Economic Competitiveness: To avoid resulting in stranded infrastructure, AIGFs should outline a competitive business with credible demand pathways, potentially requiring a shift towards a service-based multi-client operating model if anchor customers with large compute demands don't materialise (Hess & Sieker, 2025). AIGFs alone likely cannot provide the sheer scale of compute potentially required to develop frontier models in the current paradigm (You & Owen, 2025). Instead, their design should account for configurations compatible with alternative research approaches where Europe may hold greater comparative advantage, and/or developing more efficient models that lag slightly behind the frontier. Further, given the uncertainty surrounding future AI compute demand, AIGF design and location should preserve flexibility—enabling scaling if demand surges.

Resilience and Preparedness: Improving interoperability not only between AIGFs but also between AIGFs and AI Factories or other private EU clouds strengthens Resilience and Preparedness, as workloads might be re-allocated swiftly if one facility experiences outages. Mitigating strategic vulnerabilities requires not only the availability of sovereign compute but also the active migration of sensitive European workloads through preferential access conditions and targeted economic incentives.

Security and Defence: Relatedly, as compute becomes increasingly relevant to national security, AIGFs should incorporate appropriate security standards from the outset to strengthen security and defence (Grunewald & Gershovich, 2025). To manage Economic Competitiveness trade-offs like increased costs and reduced performance that accompany high security postures, a tiered security approach is advisable (Kothuri et al., 2025). This could include either designating specific AIGFs for sensitive workloads at stringent security levels (e.g. Nevo et al. (2024) SL4–5) while others operate under lighter requirements, or implementing tiered security within individual facilities.

European Values and Identity: The European Commission could mandate that a defined share of AIGF capacity be provisioned with confidential computing environments supporting attestation and logging, enabling cryptographic evidence for governance-relevant claims (e.g. that a specific model evaluation was executed under certified conditions). In parallel, a quantified share of public capacity could be reserved and prioritised for public-interest applications such as healthcare or education, ensuring AIGFs actively advance European Values and Identity.

Foreign Relations and Perceptions: Planned AIGF capacity of 100,000 H100-equivalents each may be insufficient to generate the geopolitical leverage the EU needs to shape international AI governance on its own terms. A better capacity benchmark might be a relative target of 15% of worldwide AI compute by 2030 (Shamir & Negele, 2025). Similarly, configuring parts of AIGF capacity for compute-sharing with strategic partners could position the EU as a values-aligned alternative to US or Chinese digital infrastructure dependencies. Lastly, designing AIGFs to be interoperable from the outset might ease any later multinational compute and frontier AI development pooling efforts.

# 6. Conclusion

**Frontier AI is not a narrow challenge but a general-purpose technology reshaping every pillar of European sovereignty—from economic competitiveness and defence to democratic governance and global standing.** The title of this report is a warning: the most consequential dependencies are not always imposed by force. They can arrive as gifts—through infrastructure investments, model access agreements, or technological partnerships with short-term upsides that amplify foreign interests and hinder sovereignty structurally. Europe has shown it can collectively respond to acute crises. The challenge now is to translate that reactive capacity into a durable sovereignty strategy, this time under conditions of deep technological transformation, geopolitical tension, and structural interdependence. The question is no longer whether Europe needs frontier AI sovereignty, but whether it will act with sufficient ambition and coherence before others set the terms.

**In this report, we have argued that doing so requires clearer strategic analysis and decision-making.** Existing initiatives, from the AI Continent Action Plan to the InvestAI initiative or the Franco-German Frontier AI initiative, represent genuine progress. Yet a unifying vision is lacking. The sovereignty debate has too often turned into a narrow discussion on economic competitiveness, where strategic ambition is channelled through a push for deregulation or ad hoc industrial policymaking. This report pushes against this line of thinking. Starting from an understanding of sovereignty as the capacity for self-determination, we have proposed five distinct but interconnected pillars: (i) Economic Competitiveness; (ii) Resilience and Preparedness; (iii) Security and Defence; (iv) European Values and Identity; and (v) Foreign Relations and Perceptions. Understanding the trade-offs made among these pillars is crucial to understand where strategic advantages (or threats) lie.

**The unified framework introduced in this report provides the analytical foundation for strategy-setting toward frontier AI sovereignty.** By mapping policy interventions across a granular decomposition of the frontier AI stack—five layers, 26 components, 29 sub-components—and against the five sovereignty pillars, it offers policymakers a common language for assessing interventions that current policy debates lack. Our contribution is twofold: (1) making trade-offs across sovereignty pillars explicit where they are currently implicit; and (2) surfacing the blind spots and leverage points in the frontier AI stack.

**While the framework's most consequential application lies ahead, helping Europe achieve meaningful frontier AI sovereignty, it is designed for dual application.** Looking backwards, it allows existing EU initiatives to be assessed against their actual contribution to each sovereignty pillar, as demonstrated in the analysis of planned AI Gigafactories in Section 5.4. Looking forward, it provides a structured basis for deriving targeted interventions: identifying which components of the stack demand priority investment, where leverage is most notable, or how different policy interventions interact across pillars. Frontier AI sovereignty is thus not treated as an end in itself but as enabling sovereignty overall, the means through which Europe can pursue a self-determined path through this technological transformation.

**The analytical foundations laid here will need to be stress-tested against a rapidly evolving frontier, as capabilities advance, geopolitical alignments shift, and the EU's own strategic priorities continue to change.** As a tool for policymakers and strategists navigating the frontier AI transition, its success will depend on its usefulness and adaptability. The KPIs proposed for each sovereignty pillar are starting points, not settled benchmarks. And the framework's full potential will only be released through forward-looking application: translating its diagnostics into concrete interventions aiming to address the most critical dependencies first. In this context, the benefit of deals and dependencies in frontier AI must be assessed not only for what they offer in the short run, but for what they entail in the future. Europe should identify the key strategic levers and remain vigilant about the gifts the frontier AI industry offers in light of its impact on sovereignty.

# 7. Acknowledgements

We want to thank the following people for their comments on earlier drafts: Kathrin Gardhouse, Sven Herrmann, Chloé Touzet, Bruno Galizzi, and Jimmy Farrell. We also want to thank Liza Adhiambo for copyediting and graphics design support, Amber Ace for proofreading the manuscript, and Eloise Dunn for assistance with the online publication of this report.

# Appendix

## A.1. EU AI Policy Initiatives Database

| # | Policy Domain | Initiative | Strategic Horizon | Source |
|---|---|---|---|---|
| 1 | Computing Infrastructure | EuroHPC Supercomputers | Operational | COM165 |
| 2 | Computing Infrastructure | AI Factory Antennas | Operational | COM165 |
| 3 | Data | Alliance for Language Technologies (ALT-EDIC) | Operational | COM165 |
| 4 | Data | Simpl (Shared Cloud Software) | Operational | COM165 |
| 5 | Data | European Open Science Cloud | Operational | COM165 |
| 6 | Data | Copernicus | Operational | COM165 |
| 7 | Research and Projects | GenAI4EU | Operational | COM165 |
| 8 | Research and Projects | AI-powered Advanced Screening Centres (Healthcare) | Operational | COM723 |
| 9 | Research and Projects | European Network of Expertise on AI Deployment in Healthcare | Operational | COM723 |
| 10 | Research and Projects | AI Drug Discovery Challenge | Operational | COM723 |
| 11 | Research and Projects | Streamline Market Entry of Medical Devices | Operational | COM723 |
| 12 | Research and Projects | Catalyst for Uptake of European Robotics | Operational | COM723 |
| 13 | Research and Projects | Acceleration Pipelines for Manufacturing | Operational | COM723 |
| 14 | Research and Projects | AI-enabled Situational Awareness and C2 (Command and Control) Capacities | Operational | COM723 |
| 15 | Research and Projects | Autonomous Drive Ambition Cities | Operational | COM723 |
| 16 | Research and Projects | European Telco AI Platform | Operational | COM723 |
| 17 | Research and Projects | AI Models for Energy System | Operational | COM723 |
| 18 | Research and Projects | Open Source AI Earth-system Frontier Model (DestinE) | Operational | COM723 |
| 19 | Research and Projects | Agri-Food AI platform | Operational | COM723 |
| 20 | Research and Projects | Micro-Studios for AI-Enhanced Virtual Production | Operational | COM723 |
| 21 | Research and Projects | Pan-European Multilingual AI Platforms for media | Operational | COM723 |

| # | Policy Domain | Initiative | Strategic Horizon | Source |
|---|---|---|---|---|
| 22 | Research and Projects | AI Toolbox for Public Administrations | Operational | COM723 |
| 23 | Research and Projects | Generative AI for Public Administrations | Operational | COM723 |
| 24 | Research and Projects | Materials Commons for Europe | Operational | COM724 |
| 25 | Skills | AI Skills Academy | Operational | COM165 |
| 26 | Skills | AI Fellowship Schemes | Operational | COM165 |
| 27 | Skills | Advanced Digital Skills Competitions | Operational | COM165 |
| 28 | Skills | 'MSCA Choose Europe' Scheme | Operational | COM165 |
| 29 | Skills | Sectoral AI Literacy Trainings | Operational | COM724 |
| 30 | Skills | AI for Business Executive Programmes | Operational | COM724 |
| 31 | Skills | AI Entrepreneurs Lab | Operational | COM724 |
| 32 | Skills | Talent Partnership (ICT) | Operational | COM165 |
| 33 | Skills | Multipurpose Legal Gateway Offices | Operational | COM165 |
| 34 | Skills | RAISE Doctoral Networks on AI in Science | Operational | COM724 |
| 35 | Skills | MSCA AI Projects | Operational | COM724 |
| 36 | Skills | ResearchComp Update | Operational | COM724 |
| 37 | Skills | Mutual Learning Exercise (MLE) on AI in Science | Operational | COM724 |
| 38 | Regulation and Standards | AI Act Service Desk | Operational | COM165 |
| 39 | Regulation and Standards | General-Purpose AI Code of Practice | Operational | COM165 |
| 40 | Regulation and Standards | AI Regulatory Sandboxes | Operational | COM165 |
| 41 | Regulation and Standards | Standardisation (AI Act Support) | Operational | COM165 |
| 42 | Regulation and Standards | Data Act provisions on Cloud Switching | Operational | COM165 |
| 43 | Regulation and Standards | Standardisation Request on AI Energy Consumption | Operational | COM165 |
| 44 | Regulation and Standards | Living Guidelines on Responsible Use of Generative AI in Research | Operational | COM724 |
| 45 | Regulation and Standards | AI Act Research Exemption Analysis | Operational | COM724 |

| # | Policy Domain | Initiative | Strategic Horizon | Source |
|---|---|---|---|---|
| 46 | Funding and Investment | EIC Fund | Operational | COM165 |
| 47 | Funding and Investment | InvestEU Guarantee | Operational | COM165 |
| 48 | Funding and Investment | ERC Grants for AI in Science | Operational | COM724 |
| 49 | Funding and Investment | EIC AI Support | Operational | COM724 |
| 50 | Market Structure and Ecosyst... | European Digital Innovation Hubs (EDIHs) | Operational | COM165 |
| 51 | Market Structure and Ecosyst... | Testing and Experimentation Facilities for AI (TEFs) | Operational | COM165 |
| 52 | Market Structure and Ecosyst... | Apply AI Alliance | Operational | COM723 |
| 53 | Market Structure and Ecosyst... | AI Observatory | Operational | COM723 |
| 54 | Market Structure and Ecosyst... | JRC Scientific AI Hub | Operational | COM724 |
| 55 | Market Structure and Ecosyst... | Public Sector Use Mandates (GenAI4EU) | Operational | COM165 |
| 56 | Market Structure and Ecosyst... | AI Gigafactories | Strategic | COM165 |
| 57 | Computing Infrastructure | Cloud and AI Development Act | Strategic | COM165 |
| 58 | Regulation and Standards | AI Act Implementation | Strategic | COM165 |
| 59 | Regulation and Standards | European Interoperability Framework | Strategic | COM723 |
| 60 | Regulation and Standards | Advanced Materials Act | Strategic | COM724 |
| 61 | Regulation and Standards | Biotech Act | Strategic | COM724 |
| 62 | Regulation and Standards | Strategy for Research and Technology Infrastructure | Strategic | COM724 |
| 63 | Regulation and Standards | Frontier AI Initiative | Strategic | COM723 |
| 64 | Regulation and Standards | Dedicated Secured Compute for Defence and Space AI | Strategic | COM723 |
| 65 | Funding and Investment | InvestAI | Strategic | COM165 |
| 66 | Funding and Investment | Scaleup Europe Fund (TechEU) | Strategic | COM165 |
| 67 | Funding and Investment | Horizon Europe Investment Agenda | Strategic | COM724 |
| 68 | Funding and Investment | Cohesion Policy (ERDF) | Strategic | COM724 |
| 69 | Market Structure and Ecosyst... | AI Board | Strategic | COM723 |
| 70 | Computing Infrastructure | AI Factories | Tactical | COM165 |
| 71 | Data | Data Labs | Tactical | COM165 |
| 72 | Data | European Health Data Space | Tactical | COM165 |

| # | Policy Domain | Initiative | Strategic Horizon | Source |
|---|---|---|---|---|
| 73 | Regulation and Standards | EU Startup and Scaleup Strategy | Tactical | COM724 |
| 74 | Funding and Investment | Private Sector Pledging Campaign | Tactical | COM724 |
| 75 | Market Structure and Ecosyst... | Scientific Panel on AI | Tactical | COM724 |
| 76 | Market Structure and Ecosyst... | RAISE Thematic Networks of Excellence | Tactical | COM724 |
| 77 | Market Structure and Ecosyst... | RAISE Network of Frontier AI Labs | Tactical | COM724 |
| 78 | Market Structure and Ecosyst... | AI in Science Summit | Tactical | COM724 |
| 79 | Market Structure and Ecosyst... | ERA Action on AI in Science | Tactical | COM724 |
| 80 | Data | Quality Data for AI Initiative | Tactical | COM835 |
| 81 | Data | Common European Agricultural Data Space | Tactical | COM835 |
| 82 | Data | European Mobility Data Space | Tactical | COM835 |
| 83 | Data | European Energy Data Space | Tactical | COM835 |
| 84 | Data | European Media Data Space | Tactical | COM835 |
| 85 | Data | European Legal Data Space | Tactical | COM835 |
| 86 | Data | European Green Deal Data Space | Tactical | COM835 |
| 87 | Data | European Defence Data Space | Tactical | COM835 |
| 88 | Data | European Common Language Data Space | Tactical | COM835 |
| 89 | Regulation and Standards | European Data Quality Standard | Operational | COM835 |
| 90 | Regulation and Standards | Annotation and Labelling Practices/Standards | Operational | COM835 |
| 91 | Procurement and Demand | Anti-Data-Leakage Toolbox | Tactical | COM835 |
| 92 | Skills | Data Act Legal Helpdesk | Tactical | COM835 |

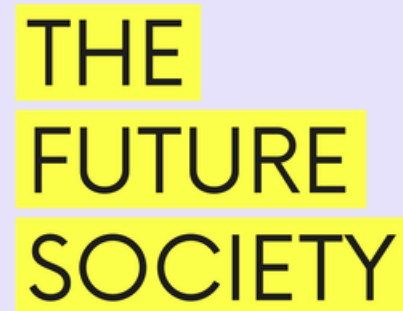

## Contact Us

**GENERAL** info@thefuturesociety.org
**PRESS** press@thefuturesociety.org

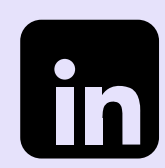

**THE FUTURE SOCIETY**
867 Boylston Street
Boston, MA, 02116
United States

www.thefuturesociety.org

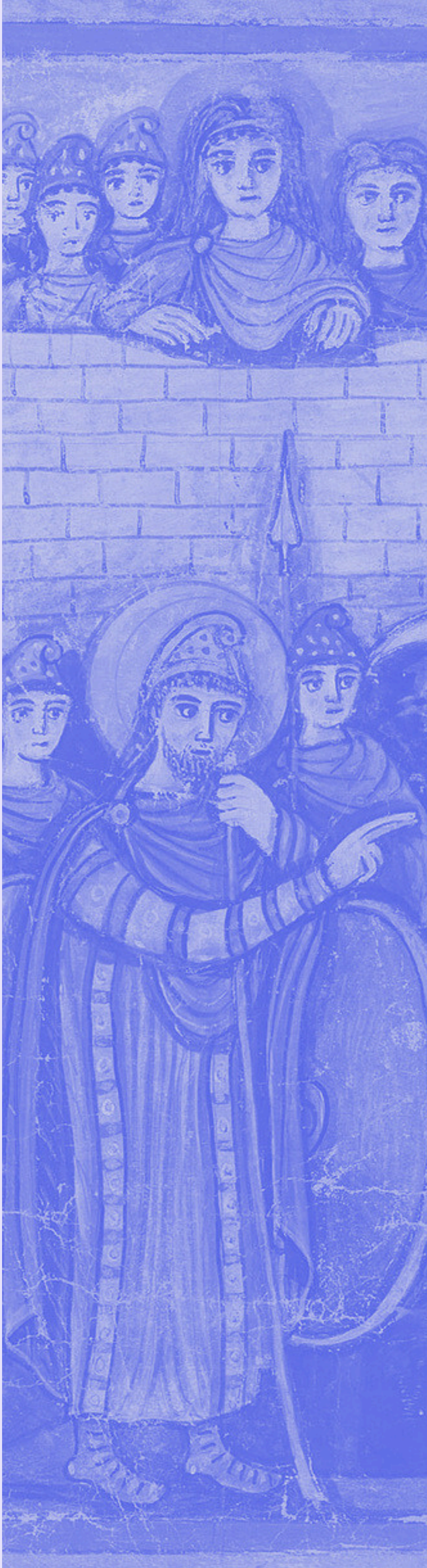